\documentclass{aastex}
\usepackage{emulateapj5,times,mathptm}
\journalinfo{The Astronomical Journal, 2003 June, astro-ph/0302558}
\slugcomment{Received 2003 February 05; accepted 2003 February 25}

\shorttitle{
X-RAY EMISSION FROM $z>4.7$ QUASARS}
\shortauthors{VIGNALI ET AL.}

\newcommand{\ltsima}{$\; \buildrel < \over \sim \;$}
\newcommand{\simlt}{\lower.5ex\hbox{\ltsima}}
\newcommand{\gtsima}{$\; \buildrel > \over \sim \;$}
\newcommand{\simgt}{\lower.5ex\hbox{\gtsima}}

\newcommand{\cgs}{ ${\rm erg~cm}^{-2}~{\rm s}^{-1}$}

\def\sun{\hbox{$\odot$}}
\def\earth{\hbox{$\oplus$}}
\def\lesssim{\mathrel{\hbox{\rlap{\hbox{\lower4pt\hbox{$\sim$}}}\hbox{$<$}}}}
\def\gtrsim{\mathrel{\hbox{\rlap{\hbox{\lower4pt\hbox{$\sim$}}}\hbox{$>$}}}}

\def\arcdeg{\hbox{$^\circ$}}
\def\arcmin{\hbox{$^\prime$}}
\def\arcsec{\hbox{$^{\prime\prime}$}}

\def\aox{$\alpha_{\rm ox}$}

\def\lumh{\rm erg s$^{-1}$ Hz$^{-1}$}
\def\ab1450{$AB_{1450(1+z)}$}

\def\xray{\hbox{X-ray}}
\def\mb{$M_{\rm B}$~}
\def\asca{{\it ASCA\/}}
\def\chandra{{\it Chandra\/}}
\def\conx{{\it Constellation-X\/}}
\def\genx{{\it Generation-X\/}}

\def\heao1{{\it HEAO-1\/}}

\def\rosat{{\it ROSAT\/}}

\def\sax{{\it BeppoSAX\/}}
\def\xeus{{\it XEUS\/}}
\def\xmm{{XMM-{\it Newton\/}}}

\begin{document}

\title{\chandra\ and \xmm\ observations of the first quasars: X-rays from the age of cosmic enlightenment}

\author{
C. Vignali,\altaffilmark{1} 
W.~N. Brandt,\altaffilmark{1} 
D.~P. Schneider,\altaffilmark{1} 
S.~F. Anderson,\altaffilmark{2}
X. Fan,\altaffilmark{3}
J.~E. Gunn,\altaffilmark{4} 
S. Kaspi,\altaffilmark{5}
G.~T. Richards,\altaffilmark{4}
and Michael~A. Strauss\altaffilmark{4} 
}
\altaffiltext{1}{Department of Astronomy and Astrophysics, The Pennsylvania State University, 
525 Davey Laboratory, University Park, PA 16802, USA; 
chris@astro.psu.edu, niel@astro.psu.edu, and dps@astro.psu.edu.}
\altaffiltext{2}{Astronomy Department, University of Washington, Box 351580, Seattle, WA 98195-1580, USA; 
anderson@astro.washington.edu.}
\altaffiltext{3}{Steward Observatory, The University of Arizona, 933 North Cherry Avenue, Tucson, AZ 85721, USA; 
fan@as.arizona.edu.}
\altaffiltext{4}{Princeton University Observatory, Peyton Hall, Princeton, NJ 08544-1001, USA; 
jeg@astro.princeton.edu, gtr@astro.princeton.edu, and strauss@astro.princeton.edu.}
\altaffiltext{5}{School of Physics and Astronomy, Raymond and Beverly Sackler Faculty of 
Exact Sciences, Tel-Aviv University, Tel-Aviv 69978, Israel; 
shai@wise.tau.ac.il.}

\begin{abstract}

We report on \chandra\ and \xmm\ observations of a sample of 13 quasars at \hbox{$z\approx$~4.7--5.4}
mostly taken from the Sloan Digital Sky Survey (SDSS). The present sample complements previous \xray\ 
studies of $z\ge4$ quasars, 
in which the majority of the objects are optically more luminous and at lower redshifts. 
All but two of our quasars have been detected in the \xray\ band, 
thus doubling the number of $z\ge4.8$ \xray\ detected quasars. 
The two non-detections are likely to be due to a short exposure time 
(SDSSp~J033829.31$+$002156.3) and to the presence of intrinsic absorption 
(SDSSp~J173744.87$+$582829.5). 
We confirm and extend to the highest redshifts the presence of a correlation between 
\ab1450\ magnitude and soft \xray\ flux for $z\ge4$ quasars, 
and the presence of a steeper optical-to-X-ray spectral energy distribution 
(parameterized by \aox) for high-luminosity, high-redshift quasars 
than for lower-luminosity, lower-redshift quasars. 
The second effect is likely due to the known anti-correlation between 
\aox\ and rest-frame 2500~\AA\ luminosity, 
whose significance is confirmed via partial correlation analysis. 
The joint \hbox{$\approx$~2.5--36~keV} rest-frame spectrum of the $z>4.8$ SDSS quasars 
observed thus far by \chandra\ 
is well parameterized by a power-law with photon index $\Gamma$=1.84$^{+0.31}_{-0.30}$; 
this photon index is consistent with those of \hbox{$z\approx$~0--3} quasars and 
that obtained from joint spectral fitting of \hbox{$z\approx$~4.1--4.5} optically luminous 
Palomar Digital Sky Survey quasars. No evidence for widespread intrinsic \xray\ absorption has been found 
(\hbox{$N_{\rm H}\simlt4.0\times10^{22}$~cm$^{-2}$} on average at 90\% confidence). 
We also obtained Hobby-Eberly Telescope (HET) photometric observations for eight of our target quasars. 
None of these shows significant ($>30\%$) optical variability over the 
time interval of a few years (in the observed frame) between the SDSS and HET observations.

\end{abstract}

\section{Introduction}

One of the main themes in modern astronomy is the study of the earliest massive objects to 
form in the Universe, now known as far back as the end of the 
reionization epoch (e.g., Rees 1999; Loeb \& Barkana 2001; Fan et al. 2002; Hu et al. 2002). 
Quasars at $z\simgt4$ can be studied in detail; 
because of their large broad-band luminosities, they are typically detectable at most wavelengths 
[e.g., 
in the radio, Schmidt et al. 1995; Stern et al. 2000; 
in the millimeter, Omont et al. 2001; Carilli et al. 2001; 
in the sub-millimeter, McMahon et al. 1999; Isaak et al. 2002; 
in the near-infrared, Barkhouse \& Hall 2001; Pentericci et al. 2003; 
in the optical, Schneider, Schmidt, \& Gunn 1989; Fan et al. 2001; 
in the \xray\ band, Kaspi, Brandt, \& Schneider 2000 (hereafter KBS00); 
Vignali et al. 2001, 2003 (hereafter V01a, V03); Bechtold et al. 2003]. 
Ground-based, large-area optical surveys such as the Sloan Digital Sky Survey (SDSS; York et al. 2000), the 
Automatic Plate Measuring facility survey (BRI; Irwin, McMahon, \& Hazard 1991), and the Palomar 
Digital Sky Survey (PSS; Djorgovski et al. 1998) 
have discovered more than 300 quasars at $z\ge4$ thus far up to $z=6.43$ 
(Fan et al. 2003).\footnote{See http://www.astro.caltech.edu/$\sim$george/z4.qsos 
for a listing of known high-redshift quasars.} 

To extend our knowledge of the \xray\ properties of $z\ge4$ quasars, 
we have started a program to observe, with \chandra\ and \xmm, both the optically most luminous 
(\mb$\approx$~$-$28.4 to $-$30.2) 
\hbox{$z\approx$~4.1--4.5} PSS/BRI quasars (e.g., V03) and the higher-redshift, optically fainter SDSS quasars 
(V01a; Brandt et al. 2002; see Brandt et al. 2003 for a recent review). 
Selecting $z\ge4$ quasar samples for \xray\ observations in two 
different regions of the luminosity-redshift parameter space allows study of 
any dependence of the \xray\ properties upon the optical selection criteria. 
In particular, it is possible to test for dependencies of $z\ge4$ quasar \xray\ continua 
on optical luminosity or redshift. 

Here we report the results obtained from 11 \chandra\ observations and one \xmm\ observation of 
a sample of 13 high-redshift \hbox{($z\approx$~4.7--5.4)}, moderately luminous 
(\mb$\approx$~$-$26.1 to $-$28.5) quasars, 
mostly discovered by the SDSS and published in Anderson et al. (2001). 
One of these quasars (RD~657; Djorgovski et al. 2003) 
was serendipitously observed and detected in the \xmm\ observation 
(i.e., there were two objects in one field).  
The only target quasar not coming from the SDSS data is BR~B0305$-$4957 
(Storrie-Lombardi et al. 2001; Peroux et al. 2001). 
Given that the accretion rate (relative to the Eddington rate) of quasars may increase 
with redshift (e.g., Kauffmann \& Haehnelt 2000), the first quasars could show unusual phenomena 
such as ``trapping radius'' effects (e.g., Begelman 1979) 
or accretion-disk instabilities (e.g., Lightman \& Eardley 1974). 
Furthermore, our targets are bright enough 
to assure in most cases \xray\ detections with snapshot \hbox{(e.g., $\approx$~4--10~ks)} observations, 
given the correlation found between \ab1450\ magnitude and soft \xray\ flux (see, e.g., V01a; V03). 
The number of $z\ge4.8$ \xray\ detected quasars is approximately doubled by the objects presented here. 
As shown in Figure~1, where the \ab1450\ magnitude is plotted as a function of redshift, 
the objects presented in this paper suitably complement previous \xray\ studies of $z\ge4$ quasars. 
All but one of these quasars are likely radio-quiet (see $\S$3 for details). 
%
\figurenum{1}
\centerline{\includegraphics[angle=0,width=8.5cm]{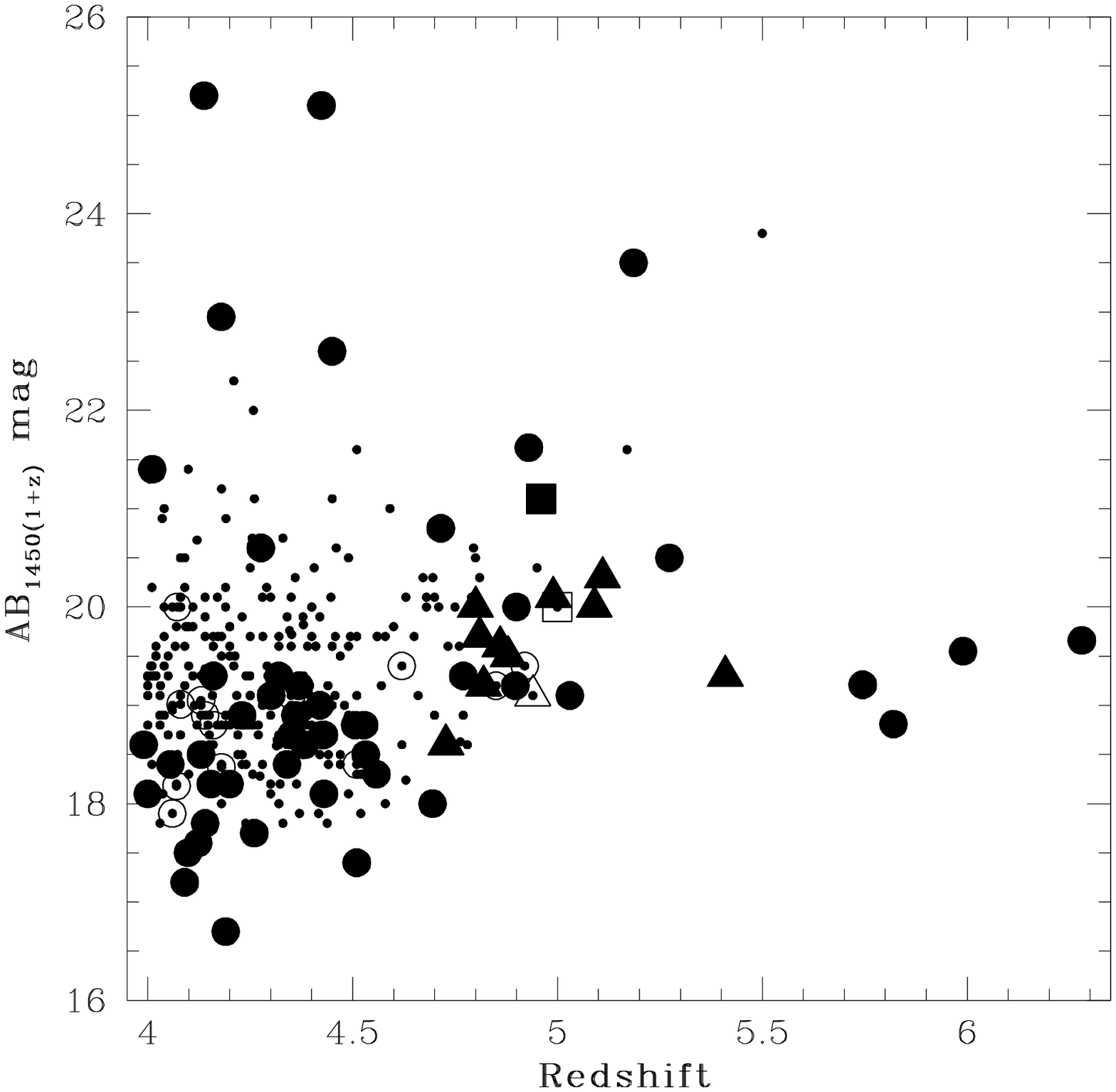}}
\figcaption{\footnotesize 
\ab1450\ versus redshift for the known $z\ge4$ 
AGNs [small dots; taken from Djorgovski as of 2002 July (see Footnote~6)]. 
The \xray\ detected (undetected) quasars presented in this paper are shown as filled (open) triangles 
(\chandra\ observations) and a filled (open) square (\xmm\ observation). 
Filled and open circles indicate $z\ge4$ \xray\ detected AGNs and tight upper limits, respectively,  
from previous observations 
[Schneider et al. 1998; Waddington et al. 1999 (note that this identification is still debated; 
L.~L. Cowie 2002, private communication); KBS00; V01a; Brandt et al. 2001, 2002; Silverman et al. 2002; 
Vignali et al. 2002 (V02); Barger et al. 2002; Bechtold et al. 2003; V03; Castander et al. 2003], 
most of which were performed by \chandra\ and \rosat.
\label{fig1}}
\centerline{}
\centerline{}

Hereafter we adopt $H_{0}$=70 km s$^{-1}$ Mpc$^{-1}$ in a $\Lambda$-cosmology 
with $\Omega_{\rm M}$=0.3 and $\Omega_{\Lambda}$=0.7 (e.g., Turner 2002).

\section{Observations and Data Reduction} 

\subsection{\chandra\ Observations}

Eleven of the 13 quasars were observed by \chandra\ during Cycle~3 
with \hbox{$\approx$~3.7--9.8~ks} observations (see Table~1; hereafter all of the quasars will be referred 
to via their abbreviated names). 
%
\begin{table*}[t]
\footnotesize
\begin{center}
{\sc TABLE 1 \\ \chandra\ and \xmm\ Observation Log}
\vskip 0.3cm
\begin{tabular}{lcccccccc}
\hline
\hline
\hskip 1.4cm Object &  & Optical & Optical & $\Delta_{\rm Opt-X}$$^{\ \rm a}$ & 
\xray & Exp.~Time$^{\ \rm b}$ & HET & \\
\hskip 1.5cm Name & $z$ & $\alpha_{2000}$ & $\delta_{2000}$ & (arcsec) & 
Obs.~Date & (ks) & Obs.~Date & Ref. \\
\hline
SDSSp~J020651.37$+$121624.4 & 4.81 & 02 06 51.4 & $+$12 16 24.4 & 0.8 & 2001 Dec 29 & 5.88 & 2001 Dec 10 & (1) \\
SDSSp~J023137.65$-$072854.5 & 5.41 & 02 31 37.6 & $-$07 28 54.5 & 0.2 & 2002 Sep 27 & 4.20 & & (1) \\
BR~B0305$-$4957 & 4.73 & 03 07 22.9 & $-$49 45 47.8 & 0.4 & 2002 Sep 22 & 3.71 & & (2,3) \\
%
SDSSp~J033829.31$+$002156.3$^{\ \rm c}$ & 5.00 & 03 38 29.3 & $+$00 21 56.3 & & 2002 Feb 22 & 5.49 & & (4) \\ 
%
RD~657$^{\ \rm c}$ & 4.96 & 03 38 30.0 & $+$00 18 39.9 & 1.7 & 2002 Feb 22 & 4.85 & & (5) \\ 
SDSSp~J075618.14$+$410408.6 & 5.09 & 07 56 18.1 & $+$41 04 08.6 & 0.2 & 2002 Feb 08 & 7.33 & 2001 Dec 09 & (1) \\
SDSSp~J075652.07$+$450258.9 & 4.80 & 07 56 52.1 & $+$45 02 58.9 & 0.5 & 2002 Oct 04 & 6.97 & 2001 Dec 10 & (1) \\
SDSSp~J091316.56$+$591921.5 & 5.11 & 09 13 16.6 & $+$59 19 21.5 & 0.7 & 2002 Mar 07 & 9.84 & 2002 Feb 07 & (1) \\
SDSSp~J094108.36$+$594725.8 & 4.82 & 09 41 08.4 & $+$59 47 25.8 & 0.2 & 2001 Oct 25 & 4.16 & 2001 Nov 16--18 & (1) \\
SDSSp~J095151.17$+$594556.2 & 4.86 & 09 51 51.2 & $+$59 45 56.2 & 0.3 & 2002 Apr 14 & 5.13 & 2002 Feb 07 & (1) \\
SDSSp~J102332.08$+$633508.1 & 4.88 & 10 23 32.1 & $+$63 35 08.1 & 0.1 & 2002 Oct 09 & 4.70 & & (1) \\
SDSSp~J173744.87$+$582829.5 & 4.94 & 17 37 44.9 & $+$58 28 29.5 & & 2002 Aug 05 & 4.62 & 2002 Jun 08 & (1) \\
SDSSp~J221644.02$+$001348.3 & 4.99 & 22 16 44.0 & $+$00 13 48.3 & 0.4 & 2002 Nov 01 & 7.41 & 2002 Jun 14 & (1) \\ 
\hline
\end{tabular}
\vskip 2pt
\parbox{6.5in}
{\small\baselineskip 9pt
\footnotesize
\indent
{\sc Note. ---} 
The optical information for the quasars presented here can be found in the papers cited in the 
reference column. 
For BR~B0305$-$4957 we use the redshift $z=4.73$ from Peroux et al. (2001) derived from 
two emission lines (Si\,{\sc iv--O\ iv]} and C\,{\sc iv}) in their high-quality optical spectrum. 
Storrie-Lombardi et al. (2001) reported $z=4.78$ (see their Table~1) and $z=4.82$ (see their Fig.~3) 
derived from a lower-quality optical spectrum with less wavelength coverage. \\
$^{\rm a}$ Distance between the optical and \xray\ positions; 
a blank entry indicates no \xray\ detection. 
$^{\rm b}$ The \chandra\ exposure time has been corrected for detector dead time. 
The \xmm\ exposure time, computed from the exposure maps (i.e., taking into account 
the decreasing effective exposure at increasing off-axis angles from the aimpoint), 
is for the observation after the background flares have been removed (see $\S$2.2). 
$^{\rm c}$ \xmm\ observation. \\
{\sc References. ---} 
Anderson et al. 2001; (2) Peroux et al. 2001; (3) Storrie-Lombardi et al. 2001; 
(4) Fan et al. 1999; (5) Djorgovski et al. 2003.
}
\end{center}
\vglue-0.9cm
\end{table*}
\normalsize
%
\begin{table*}[b]
\footnotesize
\begin{center}
{\sc TABLE 2 \\ \chandra\ sources: X-ray Counts}
\vskip 0.3cm
\begin{tabular}{lcccccc}
\hline
\hline
  & \multicolumn{4}{c}{X-ray Counts$^{\rm a}$} \\
\cline{2-5} \\
Object & [0.3--0.5~keV] & [0.5--2~keV] & [2--8~keV] & [0.5--8~keV] \\
\hline
SDSS~0206$+$1216 & $<3.0$ & {\phn}5.9$^{+3.6}_{-2.4}$ & $<4.8$ & {\phn}6.9$^{+3.8}_{-2.6}$ \\ 
SDSS~0231$-$0728 & $2.0^{+2.7}_{-1.3}$ & 22.0$^{+5.8}_{-4.7}$ & {\phn}$2.0^{+2.7}_{-1.3}$ & 23.9$^{+6.0}_{-4.9}$ \\
BR~0305$-$4957 & $<3.0$ & {\phn}$2.0^{+2.7}_{-1.3}$ & {\phn}$2.0^{+2.7}_{-1.3}$ & {\phn}4.0$^{+3.2}_{-1.9}$ \\ 
SDSS~0756$+$4104 & $2.0^{+2.7}_{-1.3}$ & 10.0$^{+4.3}_{-3.1}$ & {\phn}3.9$^{+3.2}_{-1.9}$ & 14.8$^{+4.9}_{-3.8}$ \\
SDSS~0756$+$4502 & $<4.8$ & {\phn}2.0$^{+2.7}_{-1.3}$ & $<3.0$ & {\phn}2.0$^{+2.7}_{-1.3}$ \\
SDSS~0913$+$5919 & $<4.8$ & {\phn}5.0$^{+3.4}_{-2.2}$ & $<4.8$ & {\phn}6.0$^{+3.6}_{-2.4}$ \\
SDSS~0941$+$5947 & $<3.0$ & {\phn}4.8$^{+3.4}_{-2.1}$ & $<3.0$ & {\phn}4.2$^{+3.2}_{-2.0}$ \\
SDSS~0951$+$5945 & $<3.0$ & {\phn}3.0$^{+2.9}_{-1.6}$ & $<4.8$ & {\phn}3.9$^{+3.2}_{-1.9}$ \\
SDSS~1023$+$6335 & $<3.0$ & {\phn}3.0$^{+2.9}_{-1.6}$ & $<4.8$ & {\phn}4.0$^{+3.2}_{-1.9}$ \\
SDSS~1737$+$5828 & $<3.0$ & $<3.0$ & $<3.0$ & $<3.0$ \\
SDSS~2216$+$0013 & $<4.8$ & {\phn}6.0$^{+3.6}_{-2.4}$ & $<4.8$ & {\phn}6.9$^{+3.8}_{-2.6}$ \\
\hline
\end{tabular}
\vskip 2pt
\parbox{3.7in}
{\small\baselineskip 9pt
\footnotesize
\indent
$^{\rm a}$ 
Errors on the \xray\ counts were computed according to Tables~1 and 2 of Gehrels (1986) 
and correspond to the 1$\sigma$ level; these were calculated using Poisson statistics. 
The upper limits are at the 95\% confidence level and were computed 
according to Kraft, Burrows, \& Nousek (1991).
}
\end{center}
\vglue-0.9cm
\end{table*}
\normalsize
%
All of the sources were observed with the Advanced 
CCD Imaging Spectrometer (ACIS; Garmire et al. 2003) with the S3 CCD at the aimpoint. 
Faint mode was used for the event telemetry format, 
and \asca\ grade 0, 2, 3, 4 and 6 events were used in the analysis. 
This particular grade selection is standard and appears to optimize the signal-to-background ratio 
(see $\S$6.3 of the \chandra\ Proposers' Observatory Guide). 
Source detection was carried out with {\sc wavdetect} (Freeman et al. 2002). 
We used wavelet transforms (using a Mexican hat kernel) 
with wavelet scale sizes of 1, 1.4, 2, 2.8, and 4 pixels and a 
false-positive probability threshold of 10$^{-4}$. 
Given the small number of pixels being searched due to the known source positions and the excellent 
angular resolution of \chandra, 
the probability of spurious detections is extremely low; 
most of the sources were in fact detected at a false-positive probability threshold of 10$^{-6}$. 
Typically, detections were achieved with a wavelet scale of 1.4 pixels or less. 

Source searching was performed in four energy ranges for consistency with our previous work (V01a; V03): 
the ultrasoft band \hbox{(0.3--0.5~keV)}, the soft band \hbox{(0.5--2~keV)}, the hard band \hbox{(2--8~keV)}, 
and the full band \hbox{(0.5--8~keV)}; 
in the redshift range of \hbox{$z\approx$~4.7--5.4} covered by our sample, 
the full band corresponds to the \hbox{$\approx$~2.8--51~keV} rest-frame band. 
All but one of the sources observed by \chandra\ were detected. 
The \chandra\ positions of the detected quasars were found to lie within 
\hbox{0.2\arcsec--0.8\arcsec} of their optical positions (see Table~1); 
this is consistent with the expected positional error. 
Given the \xray\ weakness of SDSS~0756$+$4502 (two full-band counts), 
we will discuss the reliability of its detection in more detail in $\S$2.1.1. 
The \chandra\ non-detection, SDSS~1737$+$5828, will be discussed further in $\S$4.1. 

The {\sc wavdetect} photometric measurements are shown in Table~2; 
they are consistent with independent measurements obtained using manual aperture photometry and a 
2\arcsec\ radius circular cell. 
Background is typically negligible in the source extraction region, except for 
SDSS~0941$+$5947, where the observation is partially affected by a background flare; 
this is taken into account in Table~2. 
The only object with enough counts to derive a useful 
power-law photon index via direct \xray\ spectral analysis is SDSS~0231$-$0728. 
Assuming only Galactic absorption, it has $\Gamma=2.4^{+0.7}_{-0.6}$ using the unbinned data and 
the \hbox{{\it C}-statistic} (Cash 1979) with {\sc xspec} Version 11.2.0 
(Arnaud 1996); errors are at the 90\% confidence level for one 
interesting parameter ($\Delta$$C=2.71$; Avni 1976; Cash 1979).\footnote{To account for the quantum efficiency decay 
of ACIS at low energies, possibly caused by molecular contamination of the ACIS filters, 
we have applied a time-dependent correction to the ACIS quantum efficiency. We generated new ancillary 
response files which have 
been used in the \xray\ spectral analyses presented throughout this paper.} 
This photon index is consistent, within the errors, with those 
of \hbox{$z\approx$~0--3} quasars in the rest-frame \hbox{2--10~keV} band (\hbox{$\Gamma\approx$~1.7--2.3}; e.g., 
George et al. 2000; Mineo et al. 2000; Reeves \& Turner 2000; Page et al. 2003). 

The soft-band images of the detected quasars (full-band image for BR~0305$-$4957), adaptively smoothed  
using the algorithm of Ebeling, White, \& Rangarajan (2003), are shown in Figure~2.

\subsubsection{\chandra\ Detection of SDSS~0756$+$4502}

Similarly to the $z=5.27$ quasar SDSS~1208$+$0010 presented in $\S$2.2 of V01a, 
SDSS~0756$+$4502 ($z=4.80$) shows only two \xray\ photons (both in the observed \hbox{0.5--2~keV band}) 
within 0.7\arcsec\ of the optical position of the source. 
The source is detected by {\sc wavdetect} using a false-positive threshold of 10$^{-4}$. 
To investigate further whether the SDSS~0756$+$4502 \xray\ detection can be considered real, we extracted a 
400$\times$400 pixel$^{2}$ ($\approx$~200$\times$200~arcsec$^{2}$) region 
centered on the optical position of the quasar, excluding the immediate vicinity of the quasar itself. 
We covered this region with 30,000 circles of 1\arcsec\ radius whose centers were randomly chosen. 
The choice of the 1\arcsec\ radius aperture appears reasonable, since 
the full-band 80\% encircled-energy radius for on-axis sources is $\approx$~0.69\arcsec\ 
(see Fig.~6.3 of the \chandra\ Proposers' Observatory Guide) and the average positional error for the other 
quasars detected by \chandra\ is $\approx$~0.5\arcsec. 
The counts obtained in each circle were averaged, giving 0.060~counts~circle$^{-1}$ in the full band and 
0.022~counts~circle$^{-1}$ in the soft band. 
The Poisson probabilities of obtaining two counts or more when 0.060 and 0.022 counts 
are expected are $\approx1.7\times10^{-3}$ and $\approx2.4\times10^{-4}$, corresponding to 
one-tailed Gaussian probabilities of \hbox{$\approx$~3.3$\sigma$} and 
\hbox{$\approx$~3.9$\sigma$}, respectively. 

We have directly verified the validity of the Poisson approximation 
for the SDSS~0756$+$4502 field using the count distribution from the 30,000-circle analysis described above. 
The results of our simulations are in reasonable agreement with the Poisson probabilities. 
Therefore, in the following, SDSS~0756$+$4502 will be considered detected by \chandra.

\subsubsection{X-ray Sources Close to the Quasars}

Following the method used in $\S$3.4 of V03, 
we have searched for possible companions for all of the objects observed by \chandra\ over a 
region of $\approx$~100\arcsec$\times$100\arcsec\ centered on the quasars 
(see Fig.~2), corresponding to a physical scale of \hbox{$\approx630\times630$~kpc$^{2}$} 
at the average redshift of our sample \hbox{($z\approx5.0$)}.
Approximately one faint serendipitous \xray\ source is found in each quasar field.  
A comparison of their \hbox{0.5--2~keV} cumulative number counts (excluding, of course, the quasars themselves) 
with those obtained from the \rosat\ deep survey of the Lockman Hole (Hasinger et al. 1998) 
and the \xmm\ moderately deep survey HELLAS2XMM (Baldi et al. 2002) 
indicates that in the proximity of the SDSS quasars there is no evidence for 
an \xray\ source overdensity. 
We found a surface density of $N(>S$)=751$^{+451}_{-301}$~deg$^{-2}$ for the 
\xray\ sources surrounding our SDSS targets (where $S$ is the soft \xray\ flux limit we reached with 
the present observations: \hbox{$\approx$~2~$\times\ 10^{-15}$~\cgs}). 
%
%
%

\begin{figure*}[!t]
\figurenum{2}
\centerline{\includegraphics[angle=0,width=\textwidth]{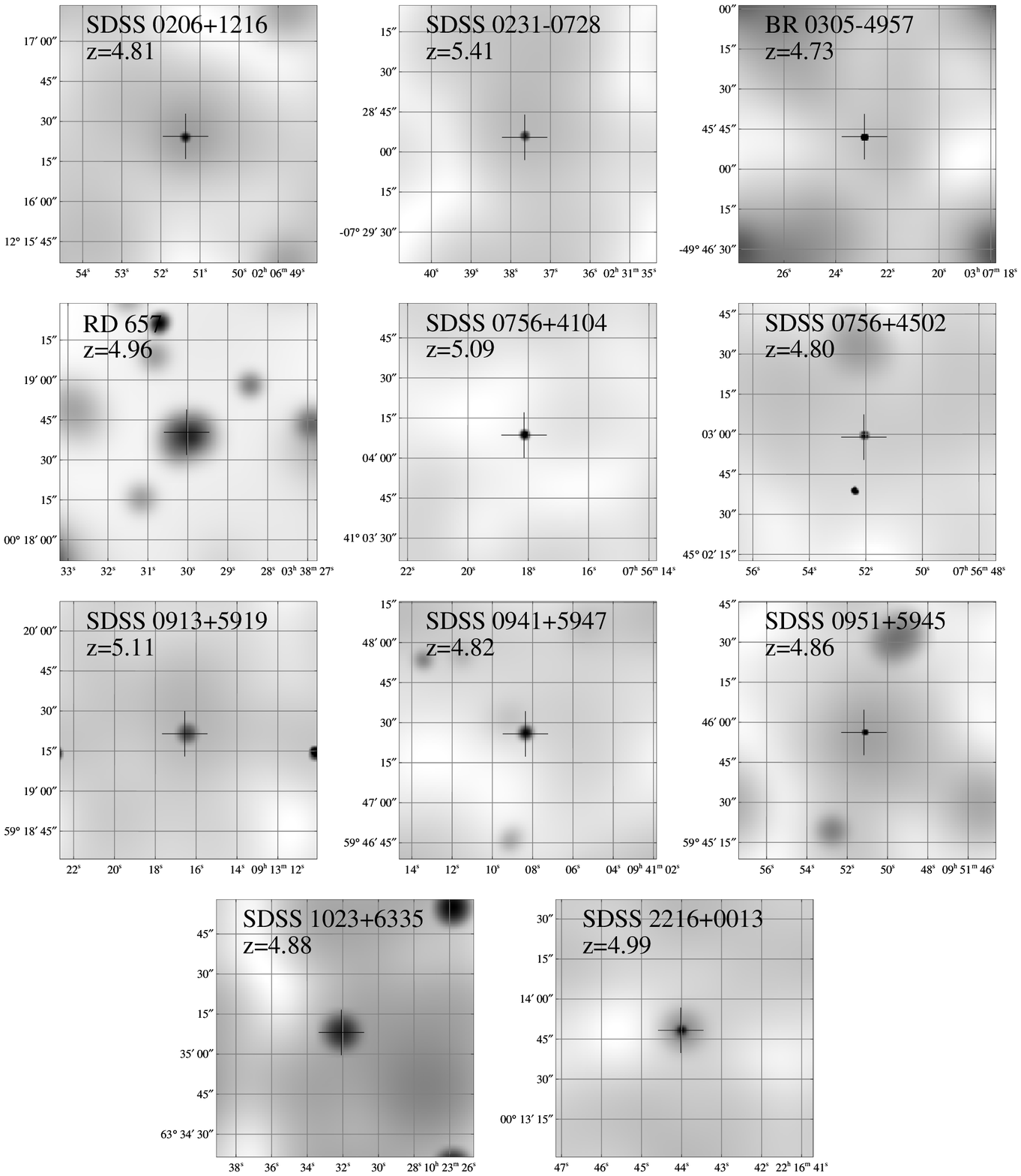}}
\vskip -3.0cm
\figcaption{\footnotesize
\chandra\ 0.5--2~keV (0.5--8~keV for BR~0305$-$4957) 
and \xmm\ 0.3--12~keV (for RD~657) 
images of the \xray\ detected quasars presented in this paper. 
In each panel, the horizontal axis shows the Right Ascension, and the vertical axis shows the 
Declination (both in J2000 coordinates). North is up, and East to the left. 
The images are 97\arcsec $\times$ 97\arcsec and have been adaptively smoothed at the 
2$\sigma$ level [except for BR~0305$-$4957, RD~657, SDSS~0951$+$5945 
(1.5$\sigma$ level), and SDSS~0756$+$4502 (1.3$\sigma$ level)] 
using the algorithm of Ebeling et al. (2003). 
Crosses mark the optical positions of the quasars. 
The image of RD~657 does not cover the position of SDSS~0338$+$0021, 
since the two quasars are separated by 196\arcsec.
\label{fig2}}
\vglue-0.2cm
\end{figure*}

\subsubsection{X-ray Extension of the Quasars}

To constrain the presence of gravitational lensing 
(e.g., Barkana \& Loeb 2000; Comerford, Haiman, \& Schaye 2002; Wyithe \& Loeb 2002), 
\xray\ jets close to the quasars (e.g., Schwartz 2002), and \xray\ scattering 
by dust in the intergalactic medium (e.g., Telis et al. 2003), 
we performed an analysis of \xray\ extension for all the quasars observed by \chandra\ 
following the procedure described in $\S$3.5 of V03. 
All of the quasars' \xray\ images are consistent with the sources being point-like. 
The separation between multiply imaged sources 
at $z\approx$~4--5 is expected to be $\approx$~1--2\arcsec\ (Barkana \& Loeb 2000). 
Typical flux ratios of multiply imaged quasars at high redshift are \hbox{$\approx$~2--8} 
(G.~Chartas 2003, private communication), being 
strongly dependent on the shape, mass, redshift, and location of the lensing system. 
Given the limited number of counts for our detected sources, we expect that only in $\approx$~30\% 
of the present sample would we be able to detect such multiple images. 
%

\subsubsection{X-ray Variability of the Quasars}

Given tentative reports of increasing quasar \xray\ variability with redshift 
(in the sense that quasars of the same \xray\ luminosity are more variable at $z>2$; 
Manners, Almaini, \& Lawrence 2002), we have searched for \xray\ variability by analyzing 
the photon arrival times in the \hbox{0.5--8~keV} band 
using the Kolmogorov-Smirnov test. No significant variability was detected from any of our quasars. 
We note, however, that the analysis of \xray\ variability for the quasars studied here is 
limited by both the short \xray\ exposures in the rest frame ($\approx$~15~min; see Table~1) 
and the low \xray\ fluxes of most of our quasars.

\subsection{\xmm\ Observation}

The SDSS~0338$+$0021 field was observed by \xmm\ for a nominal exposure time 
of 20.0~ks with the European Photon Imaging Camera (EPIC) 
pn (Str\"{u}der et al. 2001) and for 22.3~ks with the EPIC MOS (Turner et al. 2001) instruments. 
The data obtained with the Reflection Grating Spectrometer (RGS; den Herder et al. 2001), 
with a nominal exposure time of 22.8~ks, 
will not be considered in the following given the weakness of the target. 
Both the pn and MOS data were acquired in the 
``full-frame mode" using the thin filter (due to the likely weakness of the target in the \xray\ band). 
The data were reduced with the \xmm\ Science Analysis Software ({\sc sas}, Version 5.3.3) 
with the latest calibration
products; we used the {\sc epchain} and {\sc emproc} tasks to generate valid photon lists. 
We filtered the data to avoid background flares 
(due to $\approx$~100~keV protons; e.g., De Luca \& Molendi 2002) 
which affect most of this \xmm\ observation. 
To select good time intervals, 
we apply thresholds of \hbox{0.35 counts~s$^{-1}$} and \hbox{0.15 counts~s$^{-1}$} 
to pn and MOS data in the 10--12.4~keV and 10--13~keV energy bands, respectively 
(A.~Baldi 2002, private communication). In these energy bands, the counts are mostly from the background. 
The final, on-axis useful exposure time for this observation is 5.5~ks for pn and MOS. 
Both single-pixel and double-pixel events (patterns 0--4) were used 
when extracting the pn counts, while patterns 0--12 were used for the MOS. 

We performed source detection in the \hbox{0.3--2~keV}, \hbox{0.5--2~keV}, \hbox{0.5--10~keV}, 
\hbox{2--10~keV}, and \hbox{0.3--12~keV} bands using the tasks {\sc eboxdetect} and {\sc emldetect} and 
adopting a detection likelihood threshold 
of 10, corresponding to the $\approx$~4$\sigma$ detection level. 
We did not consider the energy range below 0.3~keV because 
of the enhanced background at low photon energies. 
The target, SDSS~0338$+$0021, was detected in none of the bands; its \xray\ non-detection will be 
discussed further in $\S$4.1. 

After our \xmm\ observation was assigned, Djorgovski et al. (2003) found a $z=4.96$ quasar 
(RD~657) located 196\arcsec\ from the optical position of SDSS~0338$+$0021. 
In the \xmm\ pn field-of-view we found a 4.3$\sigma$ \xray\ source detected 1.7\arcsec\ from the optical position 
of RD~657 (see Table~1) in the \hbox{0.3--12~keV} band (15.3$^{+5.0}_{-3.9}$~counts). 
We note the \xmm\ positional error is larger ($\approx$~4--5\arcsec; Metcalfe 2002) than that of \chandra; 
therefore the \xray\ and optical positions are consistent. 
Assuming the integral \hbox{0.5--2~keV} number counts of Hasinger et al. (1998) 
and an \xray\ error circle of 1.7\arcsec, 
we expect $\approx2.5\times10^{-4}$ spurious associations given the soft \xray\ flux 
of this source (3.6$\times10^{-15}$~\cgs; see Table~3). 
Therefore, we are confident that the serendipitously detected \xray\ emission 
is associated with RD~657. 
For this source we compared the 4.3$\sigma$ detection obtained by the 
{\sc sas} detection tasks with that obtained in a different way 
with {\sc SExtractor} (Bertin \& Arnouts 1996), finding similar results 
in both the number of counts and the significance of the source (3.6$\sigma$). 

Inspection of the \xray\ image at the source position confirms the \xray\ detection 
(see Fig.~2 for the 1.5$\sigma$ level adaptively smoothed image in the \hbox{0.3--12~keV} band). 
RD~657 also seems to be present in the \hbox{0.3--2~keV} adaptively smoothed image, although 
the quasar is not detected in this band by the {\sc sas} procedure described above. 
Even though there has been speculation that SDSS~0338$+$0021 and RD~657 
mark a large-scale structure, perhaps a proto-cluster at $z\approx$~5 
(Djorgovski et al. 2003), we found no clear evidence for diffuse \xray\ emission or 
an over-density of \xray\ sources around these two quasars. 
Our constraints are not tight, however, due to the short exposure time and 
relatively high background of \xmm. 
Since the background in the \xmm\ observation is higher and more complex than in the \chandra\ observations, 
for SDSS~0338$+$0021 and RD~657 we do not report the counts in individual bands 
but directly derived the main \xray\ parameters in Table~3 (see $\S$3). 
We have also analyzed the \xmm\ data with less conservative background screening choices and our main results 
are unchanged; SDSS~0338$+$0021 is not detected and RD~657 is detected.

\subsection{Hobby-Eberly Telescope Observations}

Given that quasars frequently vary in both the 
optical and \xray\ bands, one must be concerned about the reliability 
of the frequently employed \hbox{optical-to-X-ray} power-law slope whenever 
observations in the two bands are not simultaneous. 
To search for optical variability, 
eight of the 13 quasars in our sample were observed with the 9-m Hobby-Eberly Telescope 
(Ramsey et al. 1998).\footnote{HET observations were not performed for SDSS~0231$-$0728, BR~0305$-$4957, 
SDSS~0338$+$0021, RD~657, and SDSS~1023$+$6335.}  
Short (2 min; 4 min for SDSS~0941$+$5947) $I$-band images of the quasar fields were taken with 
the HET's Marcario Low-Resolution Spectrograph (LRS; Hill et al. 1998a,b; Cobos Duenas et al. 1998) 
generally within 1--2 months of the \xray\ observations 
(1--2 weeks in the rest frame; see Table~1).\footnote{The only exception is SDSS~0756$+$4502, 
observed by the HET 10 months before the \xray\ observation.} 
The HET images covered fields $\approx4\arcmin$ on a side, and using 
the published finding charts of the fields it was possible to obtain 
an approximate photometric calibration using stellar objects in the fields. 
Although it is difficult to give precise estimates given the difference 
between the LRS and SDSS bandpasses, it is clear that none of the quasars 
observed with the HET has varied 
significantly (by $\simgt 30$\%) in brightness from its published value. 
These constraints on the optical flux variability allow for at most an uncertainty of 
$\approx$~0.05 in the \aox\ determination (see $\S$3). 
The lack of large brightness variations in the sample is not surprising 
considering the relatively small 
amount of time in the rest frame (several months) 
between the SDSS photometry and the X-ray/HET observations. 

To investigate the nature of SDSS~1737$+$5828, the only quasar that was not detected in the 
\chandra\ observations, we performed a 30-min LRS spectroscopic observation. 
This was taken with the \hbox{300 line mm$^{-1}$} grating, an OG515 blocking filter, 
and a slit width of \hbox{2\arcsec} (with a seeing of $\approx$~1.6\arcsec), 
providing a spectrum in the wavelength range \hbox{5100--10200~\AA} 
at a resolving 
%
\end{multicols}
\begin{deluxetable}{lcccccccccccc}
\rotate
\tablenum{3}
\tablecolumns{13}
\tabletypesize{\footnotesize}
\tablewidth{0pt}
\tablecaption{Optical, X-ray, and Radio Properties of $z>4.7$ Quasars Observed by \chandra\ and \xmm}
\tablehead{ 
\colhead{Object} & \colhead{$N_{\rm H}$\tablenotemark{a}} & 
\colhead{$AB_{1450(1+z)}$} & \colhead{$f_{2500}$\tablenotemark{b}} & \colhead{$\log (\nu L_\nu )_{2500}$} & 
\colhead{$M_B$} & \colhead{Count~rate\tablenotemark{c}} & 
\colhead{$f_{\rm x}$\tablenotemark{d}} & \colhead{$f_{\rm 2\ keV}$\tablenotemark{e}} & 
\colhead{$\log (\nu L_\nu )_{\rm 2\ keV}$} & \colhead{$\log (L_{\rm 2-10~keV})$\tablenotemark{f}} & 
\colhead{$\alpha_{\rm ox}$\tablenotemark{g}} & \colhead{$R$\tablenotemark{h}} \\
\colhead{(1)} & \colhead{(2)} & \colhead{(3)} & \colhead{(4)} & \colhead{(5)} & \colhead{(6)} & \colhead{(7)} &  
\colhead{(8)} & \colhead{(9)} & \colhead{(10)} & \colhead{(11)} & \colhead{(12)} & \colhead{(13)}
}
\startdata
SDSS~0206$+$1216 & 6.82 & 19.7 & 0.74 & 46.6 & $-$27.5 & {\phn}1.00$^{+0.61}_{-0.40}$ & 
{\phn}4.4$^{+2.7}_{-1.8}$ & {\phn}3.83 & 44.9 & 45.1 & $-1.64\pm{0.10}$ & $<19.2$\tablenotemark{i} \\
SDSS~0231$-$0728 & 3.05 & 19.3 & 1.06 & 46.8 & $-$28.1 & {\phn}5.24$^{+1.38}_{-1.12}$ & 
21.3$^{+5.6}_{-4.6}$ & 20.36 & 45.7 & 45.9 & $-1.43^{+0.08}_{-0.07}$ & $<4.2$\tablenotemark{j} \\
BR~0305$-$4957 & 2.05 & 18.6 & 2.03 & 47.0 & $-$28.5 & {\phn}0.54$^{+0.73}_{-0.35}$ & 
{\phn}2.1$^{+2.9}_{-1.4}$ & {\phn}1.82 & 44.5 & 44.8 & $-1.94^{+0.16}_{-0.19}$ & \nodata\tablenotemark{k} \\
SDSS~0338$+$0021  & 8.11 & 20.0 & 0.56 & 46.5 & $-$27.2 & $<1.88$ & 
$<3.2$ & $<2.89$ & $<44.8$ & $<45.0$ & $<-1.65$ & $<26.0$\tablenotemark{i} \\
RD~657 & 8.05 & 21.1 & 0.20 & 46.0 & $-$26.1 & $<2.80$ & 
{\phn}3.6$^{+1.2}_{-0.9}$\tablenotemark{l} & {\phn}3.21 & 44.8 & 45.0 & $-1.46\pm{0.08}$ & $<71.3$\tablenotemark{i} \\
SDSS~0756$+$4104 & 4.76 & 20.0 & 0.56 & 46.5 & $-$27.3 & {\phn}1.36$^{+0.59}_{-0.42}$ & 
{\phn}5.7$^{+2.5}_{-1.8}$ & {\phn}5.18 & 45.1 & 45.3 & $-1.55\pm{0.09}$ & $<7.3$\tablenotemark{j} \\
SDSS~0756$+$4502 & 4.95 & 20.0 & {\phn}0.56 & 46.4 & $-$27.2 & {\phn}0.29$^{+0.39}_{-0.19}$ & 
{\phn}1.2$^{+1.7}_{-0.8}$ & {\phn}1.07 & 44.3 & 44.5 & $-1.81^{+0.16}_{-0.18}$ & $<6.3$\tablenotemark{j} \\
SDSS~0913$+$5919 & 3.81 & 20.3 & 0.42 & 46.4 & $-$27.0 & {\phn}0.51$^{+0.34}_{-0.29}$ & 
{\phn}2.1$^{+1.4}_{-0.9}$ & {\phn}1.89 & 44.6 & 44.8 & $-1.67\pm{0.11}$ & 410.8\tablenotemark{i} \\
SDSS~0941$+$5947 & 2.16 & 19.2 & 1.17 & 46.8 & $-$28.0 & {\phn}1.15$^{+0.82}_{-0.50}$ & 
{\phn}4.4$^{+3.2}_{-1.9}$ & {\phn}3.84 & 44.9 & 45.1 & $-1.72\pm{0.11}$ & $<12.2$\tablenotemark{i} \\
SDSS~0951$+$5945 & 1.44 & 19.6 & 0.81 & 46.6 & $-$27.6 & {\phn}0.59$^{+0.57}_{-0.31}$ & 
{\phn}2.2$^{+2.2}_{-1.2}$ & {\phn}1.96 & 44.6 & 44.8 & $-1.77^{+0.13}_{-0.14}$ & $<17.7$\tablenotemark{i} \\
SDSS~1023$+$6335 & 1.01 & 19.5 & 0.88 & 46.7 & $-$27.7 & {\phn}0.64$^{+0.62}_{-0.34}$ & 
{\phn}2.4$^{+2.4}_{-1.3}$ & {\phn}2.15 & 44.6 & 44.8 & $-1.77^{+0.13}_{-0.14}$ & $<16.2$\tablenotemark{i} \\
SDSS~1737$+$5828 & 3.58 & 19.1 & 1.28 & 46.8 & $-$28.1 & $<0.65$ & 
$<2.7$ & $<2.37$ & $<44.7$ & $<44.9$ & $<-1.82$ & $<3.0$\tablenotemark{j} \\
SDSS~2216$+$0013 & 4.68 & 20.1 & {\phn}0.51 & 46.4 & $-$27.1 & {\phn}0.81$^{+0.49}_{-0.32}$ & 
{\phn}3.5$^{+2.0}_{-1.4}$ & {\phn}3.09 & 44.8 & 45.0 & $-1.62\pm{0.10}$ & $<7.7$\tablenotemark{j} \\
\tableline
\enddata
\tablecomments{Luminosities are computed using $H_{0}$=70 km s$^{-1}$ Mpc$^{-1}$, 
$\Omega_{\rm M}$=0.3, and $\Omega_{\Lambda}$=0.7.}
\tablenotetext{a}{From Dickey \& Lockman (1990) in units of $10^{20}$ cm$^{-2}$ .}
\tablenotetext{b}{Rest-frame 2500~\AA\ flux density in units of $10^{-27}$ erg~cm$^{-2}$~s$^{-1}$~Hz$^{-1}$.}
\tablenotetext{c}{Observed count rate computed in the 0.5--2~keV band in units of $10^{-3}$  counts s$^{-1}$.}
\tablenotetext{d}{Galactic absorption-corrected flux in the observed 0.5--2~keV band in units 
of $10^{-15}$ erg~cm$^{-2}$~s$^{-1}$. These fluxes and the following \xray\ parameters have been corrected 
for the ACIS quantum efficiency decay at low energy.} 
\tablenotetext{e}{Rest-frame 2~keV flux density in units of $10^{-32}$ erg~cm$^{-2}$~s$^{-1}$~Hz$^{-1}$.} 
\tablenotetext{f}{Rest-frame 2--10~keV luminosity corrected for Galactic absorption 
in units of erg~s$^{-1}$.}
\tablenotetext{g}{Errors have been computed following the ``numerical method'' 
described in $\S$~1.7.3 of Lyons (1991); 
both the statistical uncertainties on the \xray\ count rates 
and the effects of the observed ranges of the \xray\ and optical continuum shapes have been taken into account 
(see the text for details; see also $\S$3 in V01a).} 
\tablenotetext{h}{Radio-loudness parameter, defined as 
$R$ = $f_{\rm 5~GHz}/f_{\rm 4400~\mbox{\scriptsize\AA}}$ (rest frame; e.g., Kellermann et al. 1989). 
The rest-frame 5~GHz flux density is computed from the observed 1.4~GHz flux density 
assuming a radio power-law slope of $\alpha=-0.8$ with $f_{\nu}\propto~\nu^{\alpha}$.}
\tablenotetext{i}{1.4~GHz flux density from NVSS (Condon et al. 1998).}
\tablenotetext{j}{1.4~GHz flux density from FIRST (Becker et al. 1995).}
\tablenotetext{k}{No radio data are available.}
\tablenotetext{l}{The soft \xray\ flux and the subsequent \xray\ parameters 
have been computed from the 0.3--12~keV counts, where the source is detected.}
\label{tab3}
\end{deluxetable}
\begin{multicols}{2}
%

\noindent 
power of \hbox{$\approx$~240--300}. 
The spectroscopic data were reduced using standard {\sc iraf} routines. 
The wavelength calibration was performed using a Ne arc lamp, 
while the flux calibration was achieved using the standard star BD~$+$17~4708, 
observed during the same night. 
The HET spectrum (black curve in Fig.~3) is discussed in $\S$4.1 along with the SDSS spectrum 
(adapted from Anderson et al. 2001; grey curve in Fig.~3). 

\figurenum{3}
\centerline{\includegraphics[angle=0,width=8.5cm]{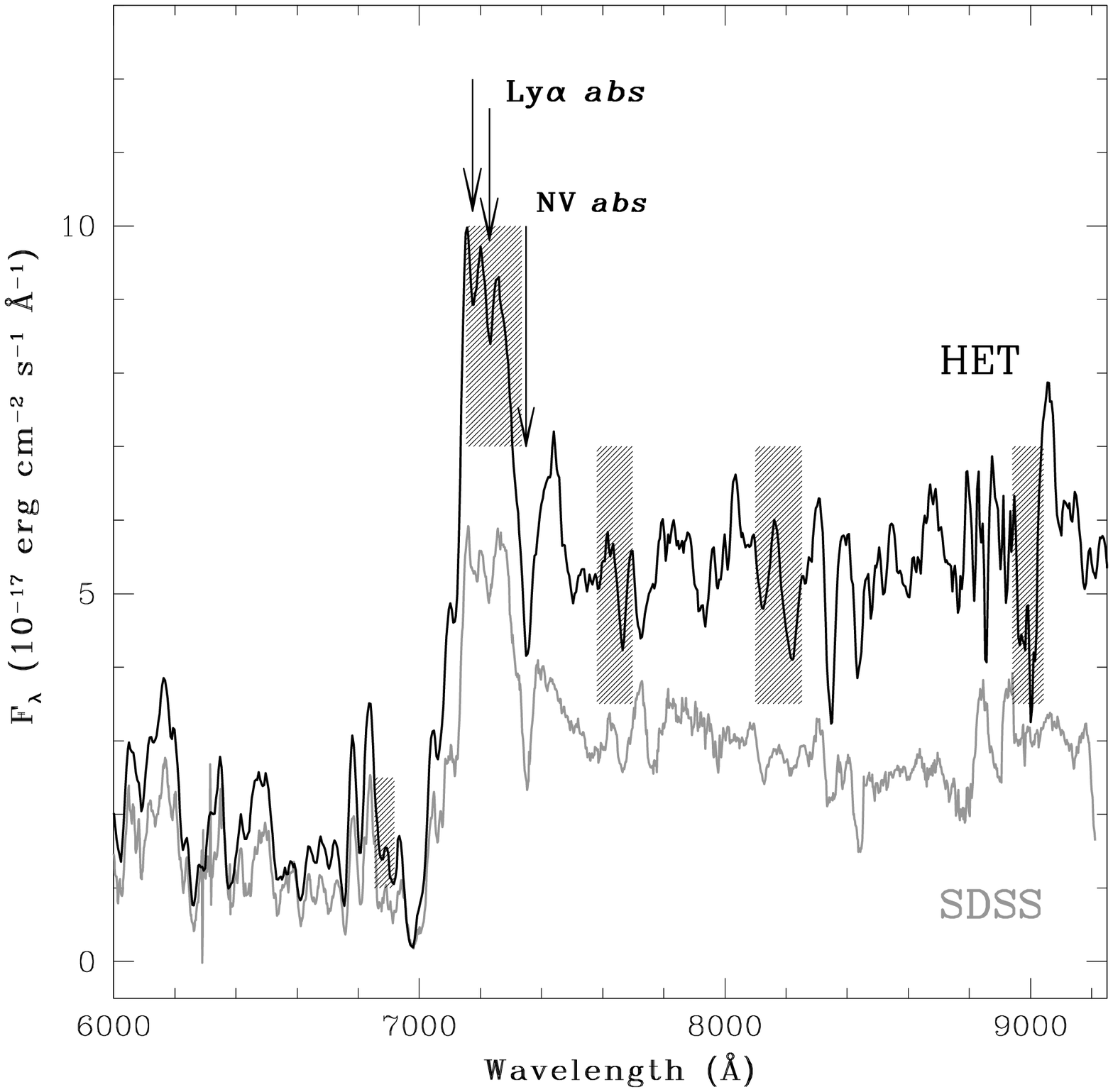}}
\figcaption{\footnotesize
Optical spectra of SDSS~1737$+$5828 obtained with the HET (30-min exposure; black curve) and by the SDSS 
(adapted from Anderson et al. 2001; grey curve). No correction for the telluric lines has been applied 
to the HET spectrum (the affected regions of the HET spectrum are shaded). 
The likely absorption features from N\,{\sc v} and Ly$\alpha$ are shown (see $\S$4.1 for details).
\label{fig3}}
\centerline{}
\centerline{}
%
A direct comparison of the 
continuum levels of the two spectra is difficult. While the uncertainty in the 
calibration of the SDSS spectrum is less than $\approx$~20\%, 
it is hard to obtain an accurate HET flux calibration mostly because of the variations of 
effective aperture produced by its optical design (in addition to problems due to slit losses).

\section{Multi-wavelength Properties of $z>4.7$ Quasars}

The principal optical, \xray, and radio properties of the quasars in our sample are shown in Table~3. 
A description is as follows: \\
{\sl Column (1)}. --- The abbreviated name of the source. \\
{\sl Column (2)}. --- The Galactic column density (from Dickey \& Lockman 1990) 
in units of \hbox{10$^{20}$ cm$^{-2}$}. \\
{\sl Column (3)}. --- The monochromatic rest-frame \ab1450\ magnitude (defined in $\S$3b of 
Schneider et al. 1989) with estimated errors $\approx$~0.1~mag. The absolute photometry is correct 
to within $\approx$~0.05 mag (Anderson et al. 2001), while the computation of the \ab1450\ magnitude 
using the photometric measurements and the composite SDSS quasar spectrum (Vanden Berk et al. 2001) 
is correct to within $\approx$~0.05--0.10 mag. \\
{\sl Columns (4) and (5)}. --- The 2500~\AA\ rest-frame flux density and luminosity. 
These were computed from the \ab1450\ magnitude assuming an optical power-law slope of 
$\alpha=-0.79$ \hbox{($S_{\nu}$ $\propto$ $\nu^{\alpha}$}; Fan et al. 2001) 
to allow direct comparison with the results presented in V01a and V03. 
The 2500~\AA\ rest-frame flux densities and luminosities are reduced by $\approx$~15\% 
when the power-law slope of the optical continuum is changed to $\alpha=-0.5$ (e.g., Vanden Berk et al. 2001; 
Schneider et al. 2001). \\
{\sl Column (6)}. --- The absolute $B$-band magnitude computed using \hbox{$\alpha=-0.79$}. 
If $\alpha=-0.5$ is adopted for the extrapolation, 
the absolute $B$-band magnitudes are fainter by $\approx$~0.35 mag. \\ 
{\sl Columns (7) and (8)}. --- The observed count rate in the \hbox{0.5--2~keV} band and the corresponding 
flux ($F_{\rm X}$), corrected for Galactic absorption. 
This flux has been calculated using {\sc pimms} (Version 3.2d; Mukai 2002) and 
a power-law model with \hbox{$\Gamma=2.0$}, as derived for samples of \hbox{$z\approx$~0--3} quasars 
(e.g., George et al. 2000; Mineo et al. 2000; Reeves \& Turner 2000; Page et al. 2003) and $z>4$ 
optically luminous, PSS quasars (V03); see also $\S$4.3.  

The \xray\ fluxes reported in this paper have been corrected for the ACIS quantum efficiency decay 
at low energy (see Footnote~7). 
For each quasar, we produced an ancillary response file (ARF) assuming 
an exponential decay for the ACIS quantum efficiency. 
From each ARF we derived an effective area that was 
used to replace the original ACIS-S effective area in {\sc pimms}. 
The \hbox{0.5--2~keV} flux correction is $\approx$~20\% for the \chandra\ observations presented here. 
All of the parameters derived from the measured soft \xray\ flux have been 
corrected. 
Following this procedure, we also corrected all of the \xray\ fluxes of the $z\ge4$ quasars 
previously observed by \chandra\ (see Table~A1 in the Appendix for an updated version of 
the relevant numbers for $z\ge4$ quasars; also see http://www.astro.psu.edu/users/niel/papers/). 

The soft \xray\ flux changes by \hbox{$\approx$~1--2\%} over a reasonable range 
in the adopted power-law photon index, \hbox{$\Gamma=$~1.7--2.3}. 
The soft \xray\ flux derived from the \hbox{0.5--2~keV} counts is generally similar 
(to within \hbox{$\approx$~5--20\%}) to that derived from the \hbox{0.5--8~keV} counts. 

For RD~657, 
the \hbox{0.5--2~keV} flux and subsequent \xray\ parameters reported below have been derived 
using the counts in the \hbox{0.3--12~keV} band, where it was detected. \\
{\sl Columns (9) and (10)}. --- The rest-frame 2~keV flux density and luminosity, computed assuming $\Gamma=2.0$. \\
{\sl Column (11)}. ---  The \hbox{2--10~keV} rest-frame luminosity, corrected for Galactic absorption. \\
{\sl Column (12)}. ---  The optical-to-X-ray power-law slope, \aox, defined as 
\begin{equation}
\alpha_{\rm ox}=\frac{\log(f_{\rm 2~keV}/f_{2500~\mbox{\rm \scriptsize\AA}})}{\log(\nu_{\rm 2~keV}/\nu_{2500~\mbox{\rm \scriptsize\AA}})}
\end{equation}
where $f_{\rm 2~keV}$ and $f_{2500~\mbox{\scriptsize \rm \AA}}$ are the rest-frame flux densities at 2~keV and 2500~\AA, 
respectively. The $\approx1\sigma$ errors on \aox\ have been computed 
following the ``numerical method'' described in $\S$~1.7.3 of Lyons (1991). 
Both the statistical uncertainties on the \xray\ count rates 
and the effects of possible changes in the \xray\ ($\Gamma\approx$~1.7--2.3) 
and optical ($\alpha\approx$~$-$0.5 to $-$0.9; Schneider et al. 2001) continuum shapes 
have been taken into account (see $\S$3 of V01a for further details). 
Changing the power-law slope of the optical continuum from $\alpha=-0.79$ to 
$\alpha=-0.5$ induces a small increase in the \aox\ values (more positive by $\approx$~0.03). \\
{\sl Column (13)}. --- The radio-loudness parameter (e.g., Kellermann et al. 1989), 
defined as \hbox{$R$=$f_{\rm 5~GHz}/f_{\rm 4400~\mbox{\scriptsize\AA}}$} (rest frame). 
The rest-frame 5~GHz flux density was computed from the FIRST (Becker, White, \& Helfand 1995) or NVSS 
(Condon et al. 1998) observed 1.4~GHz flux density assuming a radio power-law slope of $\alpha=-0.8$. 
The upper limits reported in the table are at the 3\/$\sigma$ level. 
The rest-frame 4400~\AA\ flux density was computed from the \ab1450\ magnitude assuming an 
optical power-law slope of $\alpha=-0.79$. 
Radio-loud quasars (RLQs) have $R>100$, whereas radio-quiet quasars (RQQs) are 
characterized by $R<10$ (e.g., Kellermann et al. 1989). 

SDSS~0913$+$5919 is the only radio-loud object ($R\approx$~411) in our sample and is also 
the highest redshift ($z=5.11$) RLQ thus far detected in the \xray\ band. 
For one quasar, BR~0305$-$4957, no radio data are available. 
None of the remaining quasars is a RLQ, although in some cases 
the observed-frame 1.4~GHz flux-density constraints provided by the NVSS are not sufficiently tight to 
prove that a quasar is radio quiet ($R<10$); 
at most these objects could be radio moderate (see Table~3). 
Given that {$\approx$~85--90\%} of the quasar population is radio quiet (e.g., Stern et al. 2000), 
we expect the majority of the quasars with radio-loudness upper limits of 10--100 to be radio quiet.

\section{Discussion}

Thanks mostly to \chandra\ (e.g., V01a; Brandt et al. 2002; Bechtold et al. 2003; V03) 
and \rosat\ (e.g., KBS00; V03) observations, 
we have seen a substantial growth in the data needed to study the \xray\ properties of $z\ge4$ quasars. 
At present, there are $\approx$~70 quasars at \hbox{$z\ge4$} with \xray\ 
detections.\footnote{See http://www.astro.psu.edu/users/niel/papers/highz-xray-detected.dat for a 
regularly updated compilation of \xray\ detections at $z\ge4$.}
%
\figurenum{4}
\centerline{\includegraphics[angle=0,width=8.5cm]{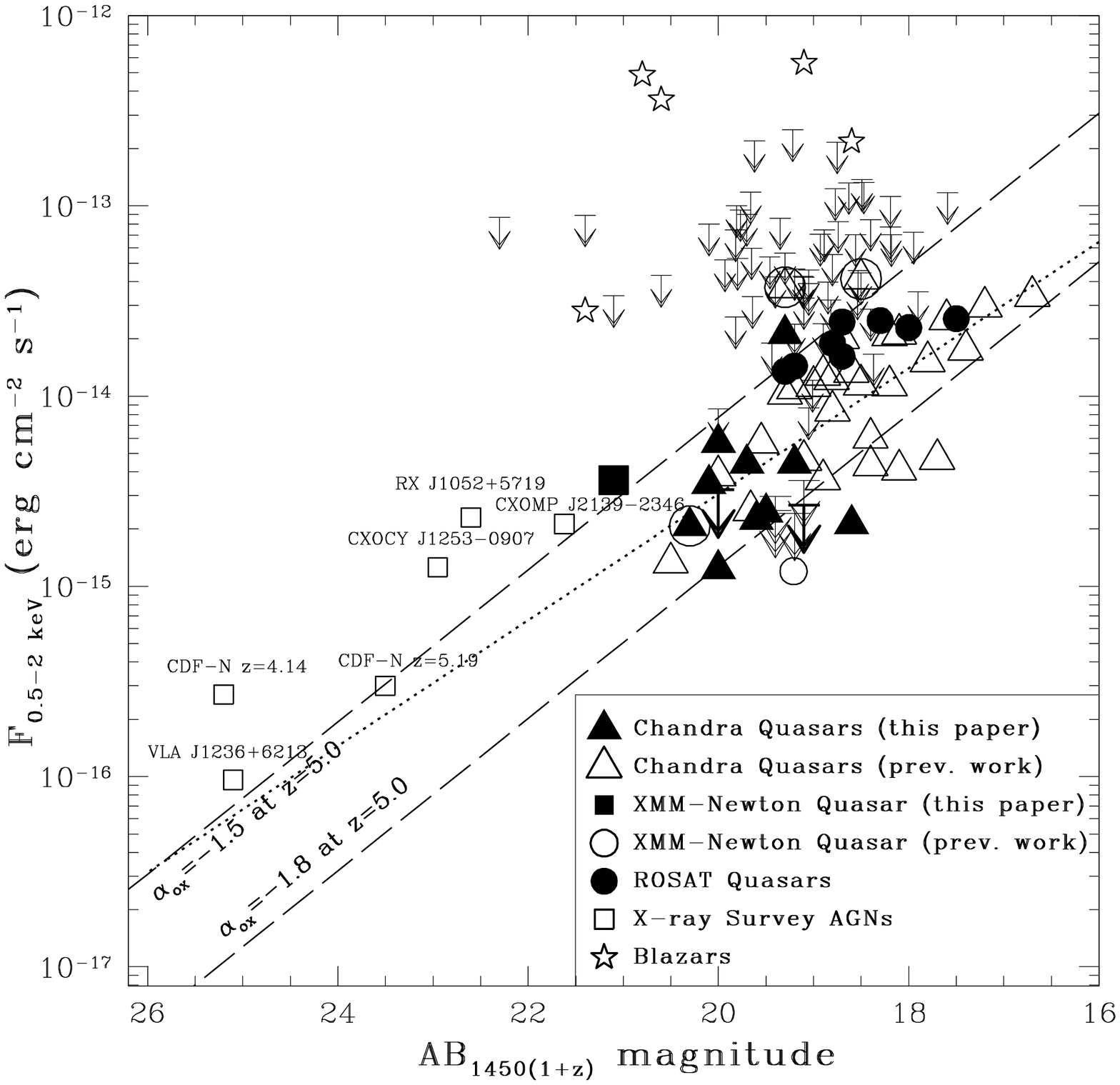}}
\figcaption{\footnotesize 
Observed-frame, Galactic absorption-corrected \hbox{0.5--2~keV} flux versus \ab1450\ 
magnitude for $z\ge4$ AGNs. Object types are shown in the key. 
In particular, the objects presented in this paper are plotted as filled triangles 
(\chandra\ detections) and a filled square (\xmm\ detection), while large, 
thick, downward-pointing arrows indicate \xray\ upper limits for the objects of our sample. 
\chandra\ RLQs are plotted as circled triangles. 
\rosat\ upper limits (small downward-pointing arrows) are at the 3$\sigma$ confidence level. 
The slanted dashed lines show the \aox=$-$1.5 and \aox=$-$1.8 loci at $z=5.0$ 
(the average redshift of the present sample). The dotted line shows the best-fit 
correlation reported in $\S$4 [Equation~(2)] for the whole sample of $z\ge4$ optically selected RQQs. 
\label{fig4}}
\centerline{}
\centerline{}
%
Figure~4 shows the observed-frame, Galactic absorption-corrected \hbox{0.5--2~keV} 
flux versus \ab1450\ magnitude for a compilation of $z\ge4$ Active Galactic Nuclei (AGNs). 
The detected objects presented in this paper are shown as filled triangles (10 \chandra\ detections) and 
a filled square (one \xmm\ detection), while large and thick downward-pointing arrows indicate the 
two quasars in our sample with \xray\ upper limits. 
Because this figure is constructed from {\it directly observed} 
physical quantities, it is robust and useful 
for planning future \xray\ observations of high-redshift quasars. 
We present a complementary discussion based on \aox, a parameter somewhat less directly tied to observation, 
in $\S$4.2. 

Quasars are well-known \xray\ emitters at $z\approx$~0--3 
(e.g., Vignali et al. 1999; Reeves \& Turner 2000; Page et al. 2003). 
Figure~4 indicates that \xray\ emission is also a universal property of $z\ge4$ quasars. 

Previous studies of $z\ge4$ quasars have shown that there is a significant correlation 
between \ab1450\ and \hbox{0.5--2~keV} flux (V01a; V03). The new \xray\ detections and upper limits 
provided by the present work significantly improve the study of this correlation at fainter optical magnitudes. 
The average \ab1450\ magnitude of the sample presented here is 19.7, whereas the average 
for the bright PSS sample of V03 is 17.9. 
The presence of this correlation indicates that the same engine 
(namely accretion onto a supermassive black hole) is powering both the ultraviolet and \xray\ emission. 
To study this correlation using all the available \xray\ detections (41) 
and upper limits (52) for $z\ge4$ optically selected RQQs 
(the majority of the objects plotted in Fig.~4), 
we have used the {\sc asurv} software package Rev~1.2 (LaValley, Isobe, \& Feigelson 1992). 
To evaluate the significance of the correlation, we used several methods available in {\sc asurv}, 
namely the Cox regression proportional hazard (Cox 1972), 
the generalized Kendall's $\tau$ (Brown, Hollander, \& Korwar 1974), and the Spearman's $\rho$ models. 
We have chosen to use only optically selected RQQs in order to have a homogeneous sample 
(see $\S$4 of Vignali et al. 2002, hereafter V02, for the results of a sample of \xray\ selected quasars; 
also see Silverman et al. 2002 and Castander et al. 2003). 
Once the Broad Absorption Line quasars (BALQSOs), which are known to be 
absorbed in the \xray\ band (e.g., Brandt, Laor, \& Wills 2000; 
Gallagher et al. 2001, 2002; Green et al. 2001; V01a), 
are excluded from the analysis (there are four in the original sample), 
we find that the \ab1450\ versus $F_{\rm X}$ correlation 
is significant at the $>99.9$\% level (all the tests above provide similar results). 
According to the EM (Estimate and Maximize) regression algorithm 
(Dempster, Laird, \& Rubin 1977), the correlation is parameterized by 
\begin{equation}
\log (F_{\rm X}/{\rm erg~cm^{-2}~s^{-1}})=[-(0.33\pm{0.05})\ AB_{1450(1+z)}-(7.89\pm{0.91})],
\end{equation}
which is plotted as a dotted line in Figure~4. 
Since \ab1450=$-2.5\ \log f_{1450~\mbox{\scriptsize\AA}} -48.6$, 
the previous equation may also be written as 
\begin{equation}
\log (F_{\rm X}/{\rm erg~cm^{-2}~s^{-1}})\ \propto \ (0.83\pm{0.13})\ \log f_{1450~\mbox{\scriptsize\AA}}.
\end{equation}

In Figure~5 the quantity $F_{\rm X,obs}$/$F_{\rm X,fit}$ 
is shown as a function of \ab1450\ magnitude 
for the $z\ge4$ optically selected RQQs observed by \chandra. 
$F_{\rm X,obs}$ represents the measured, Galactic absorption-corrected 0.5--2~keV flux, and 
$F_{\rm X,fit}$ the flux expected from the best-fit correlation above. 
The majority of the $z\ge4$ quasars observed by \chandra\ are within a factor of two of the best-fit 
correlation. The only quasar clearly showing a large difference between the measured and expected fluxes is 
SDSS~0231$-$0728 (the brightest \xray\ source in our sample). Since its \xray\ 
emission is unlikely to be contaminated by nearby \xray\ sources (see Fig.~2), 
it appears plausible that its unusually large \xray\ flux is due to an episode of variability. 
Unfortunately, the lack of an  HET observation for SDSS~0231$-$0728 does not allow us to examine this hypothesis 
further (assuming that the \xray\ and optical variability are correlated). 
%
\figurenum{5}
\centerline{\includegraphics[angle=0,width=8.5cm]{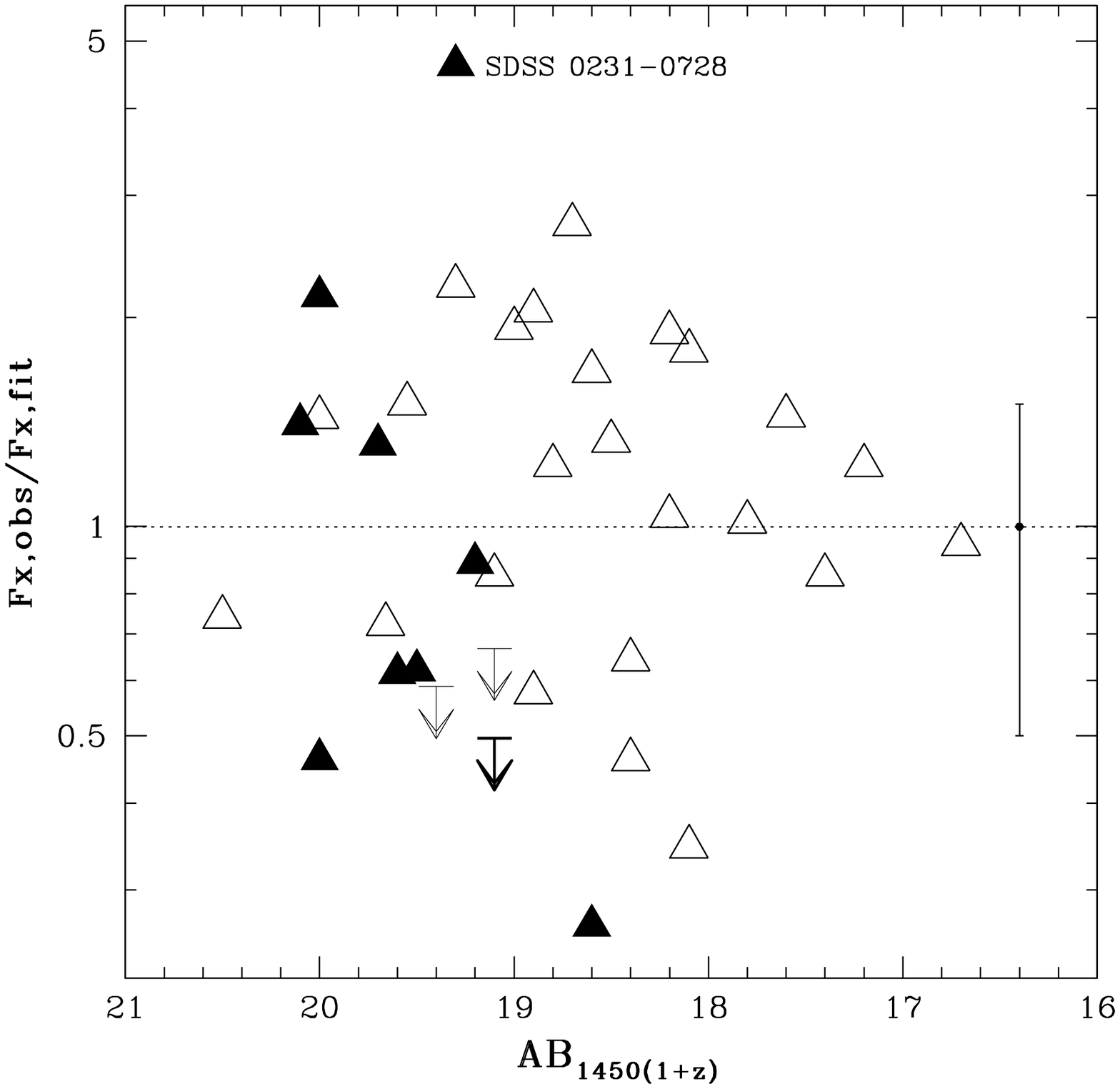}}
\figcaption{\footnotesize 
Plot of $F_{\rm X,obs}$/$F_{\rm X,fit}$ as 
a function of \ab1450\ magnitude for the $z\ge4$ optically selected RQQs observed by \chandra. 
$F_{\rm X,obs}$ represents the Galactic absorption-corrected 0.5--2~keV flux 
observed by \chandra\ and $F_{\rm X,fit}$ the flux expected on the basis of the 
best-fit correlation between the soft \xray\ flux and \ab1450\ magnitude discussed 
in $\S$4 [Equation~(2)]. 
Symbols are the same as in Figure~4. 
The dotted line indicates where the measured and expected fluxes are the same.
To avoid symbol crowding, we have only shown a representative average uncertainty 
on the flux ratios (the error bar near the middle right). 
\label{fig5}}
\centerline{}
\centerline{}
%
A constant fit to the data points of Figure~5 (including the 
associated errors, but excluding the upper limits) is statistically unacceptable ($\chi^{2}=54.6$ for 
33 degrees of freedom). Excluding SDSS~0231$-$0728 improves the quality of the fit 
($\chi^{2}=45.3$), although it remains unacceptable. 
This indicates that there is a real dispersion in the quantity plotted in Figure~5 above the statistical 
noise (see also $\S$4.2).

\subsection{X-ray Non-Detections: Likely Explanations}

All but two (SDSS~0338$+$0021 and SDSS~1737$+$5828) 
of the quasars presented here have been detected in the \xray\ band. 
The non-detection of SDSS~0338$+$0021 is probably due to the short exposure time of 
the \xmm\ observation once the flaring-background intervals have been removed. 
Assuming the best-fit parameters of the \ab1450\ versus \hbox{0.5--2~keV} flux 
correlation reported in Equation (2), 
we would expect a \hbox{0.5--2~keV} flux of $\approx3.2\times10^{-15}$~\cgs. 
Although SDSS~0338$+$0021 was observed at the aimpoint, 
the on-axis flux limit of the \xmm\ EPIC pn in a $\approx$~5.5~ks exposure (see Table~1) is 
a factor of $\approx$~2 higher than that achieved with \chandra\ in the 
same amount of time. 

No \chandra\ source counts were detected at the optical position of SDSS~1737$+$5828. 
The faint \rosat\ source reported as a possible counterpart of SDSS~1737$+$5828 by Anderson et al. (2001) 
is clearly present in the \chandra\ observation, but its \xray\ position is 
45\arcsec\ from the quasar's optical position. 
There are two possible explanations for the SDSS~1737$+$5828 non-detection: 
this quasar is either strongly absorbed or intrinsically \xray\ weak. 
If the non-detection is due to the presence of intrinsic \xray\ absorption, 
a column density of $N_{\rm H}\simgt1.2\times10^{23}$~cm$^{-2}$ is necessary 
to reproduce the \chandra\ constraints (see Tables~2 and 3) 
assuming the average \aox=$-1.71$ of the sample (see $\S$4.2) and an \xray\ photon index of $\Gamma=2.0$. 
The absorption possibility finds partial support from the published 
SDSS~1737$+$5828 optical spectrum (Anderson et al. 2001; 
see the grey curve in Fig.~3), which shows 
evidence of a strong, narrow \ion{N}{5} absorption feature at 
$z\approx$~4.92. 
This absorption feature has a fairly high formal significance in the 
SDSS spectrum, but given the importance of the intrinsic absorption 
interpretation, we wished to acquire an independent confirmation of the 
line; the new observation would also allow an investigation of any 
line variability. We therefore obtained a 
\hbox{30-min} spectroscopic observation of SDSS~1737$+$5828 with the 
HET (see $\S$2.3). The resulting spectrum (black curve in Fig.~3) confirms the 
presence of a clear narrow \ion{N}{5} absorption feature.  This line 
is unresolved (the spectral resolution is 
\hbox{$\approx$~1000~km~s$^{-1}$}) and has a blueshift of 
\hbox{$\approx$~700~km~s$^{-1}$} relative to the quasar redshift of 
4.94 reported by Anderson et al. (2001).
In both the SDSS and HET spectra 
we see evidence for narrow Ly$\alpha$ absorption spanning the 
redshift of the \ion{N}{5} absorption feature with the strongest 
absorption appearing at \hbox{$z\approx$~4.90} and \hbox{$z\approx$~4.945}. There may be 
features at similar redshifts in \ion{Si}{4} and \ion{C}{4}, but the 
signal-to-noise ratio of the spectra prevents us from unambiguously identifying these 
systems. Nevertheless, the \ion{N}{5} line alone strongly suggests 
that significant internal absorption is present in SDSS~1737$+$5828, 
since \ion{N}{5} absorption of this strength is rarely caused by 
intervening galaxies (e.g., York et al. 1991) 
and the absorption is blueshifted with respect to the emission-line redshift. 
The Ly$\alpha$ emission line appears weak; the rest-frame equivalent width 
is only $\approx$~35~\AA. 
On average, the rest-frame equivalent width of Ly$\alpha$ $+$ \ion{N}{5} in
high-redshift quasars is \hbox{$\approx$~70--80~\AA} (Schneider, Schmidt, \& Gunn 1991; Fan et al. 2001). 
The evidence from the HET and SDSS spectra strongly suggests that significant 
internal absorption is present in SDSS~1737$+$5828, but the quasar is 
clearly not a strong high-redshift BALQSO such as SDSS~0856$+$5252 
($z=4.79$; Anderson et al. 2001), SDSS~1044$-$0125 ($z=5.74$; Fan et al. 2000; 
Brandt et al. 2001; Maiolino et al. 2001; Goodrich et al. 2001), 
or SDSS~1129$-$0142 ($z=4.85$; Zheng et al. 2000).


\subsection{\aox\ Results}

Previous reports (e.g., V01a; Bechtold et al. 2003; V03) of 
steeper \aox\ indices for $z\simgt4$ quasars than for 
lower-luminosity, lower-redshift quasars are confirmed by the present analysis; see Figure~6. 
The 12 RQQs of our sample have $\langle\alpha_{\rm ox}\rangle$=$-$1.71$\pm{0.05}$ 
(the error represents the standard deviation of the mean). 
This average \aox\ value is considerably steeper 
than that obtained for, e.g., the Bright Quasar Survey (BQS; Schmidt \& Green 1983) RQQs at \hbox{$z<0.5$} 
(\hbox{$\langle\alpha_{\rm ox}\rangle$=$-$1.56$\pm{0.02}$}; see Brandt et al. 2000 and V01a). 
The sample of eight optically luminous $z>4$ PSS RQQs presented in V03 has an even steeper average \aox\ 
(\hbox{$\langle\alpha_{\rm ox}\rangle$=$-$1.77$\pm{0.03}$}).\footnote{This average value is flatter by 0.04 
than that originally reported in V03 because of the flux correction discussed in $\S$3.}
The PSS sample of V03 has 
\hbox{$\langle\log\ (l_{2500~\mbox{\rm \scriptsize\AA}}/{\rm erg~s^{-1}~Hz^{-1}}) \rangle=32.2$}, 
while the one presented here has 
\hbox{$\langle\log\ (l_{2500~\mbox{\rm \scriptsize\AA}}/{\rm erg~s^{-1}~Hz^{-1}}) \rangle=31.5$}. 
According to the known anti-correlation between \aox\ and 2500~\AA\ luminosity 
(e.g., Avni, Worrall, \& Morgan 1995; Vignali, Brandt, \& Schneider 2003, hereafter VBS03; 
also see the discussion later in this subsection) following the equation\footnote{This 
is an updated version of Equation~(2) of VBS03, taking into account the correction for the quantum 
efficiency decay of ACIS.} 
\begin{equation}
\alpha_{\rm ox}=-(0.095\pm{0.021})\times\ l_{2500~\mbox{\rm \scriptsize\AA}}+(1.295\pm{0.648}),
\end{equation}
we expect an average \aox\ of $-$1.70 for the sample of RQQs presented here, in good agreement 
with the average \aox\ value reported above. 
A summary of these results is shown in Figure~6: the \aox\ distribution for the BQS RQQs 
(top panel) is clearly different from that of the $z\ge4$ RQQs observed by \chandra\ thus far 
(bottom panel). In both \aox\ distributions, some of the quasars with the steepest \aox\ values 
are BALQSOs or mini-BALQSOs; areas with arrows indicate upper limits. 
%
\figurenum{6}
\centerline{\includegraphics[angle=0,width=8.5cm]{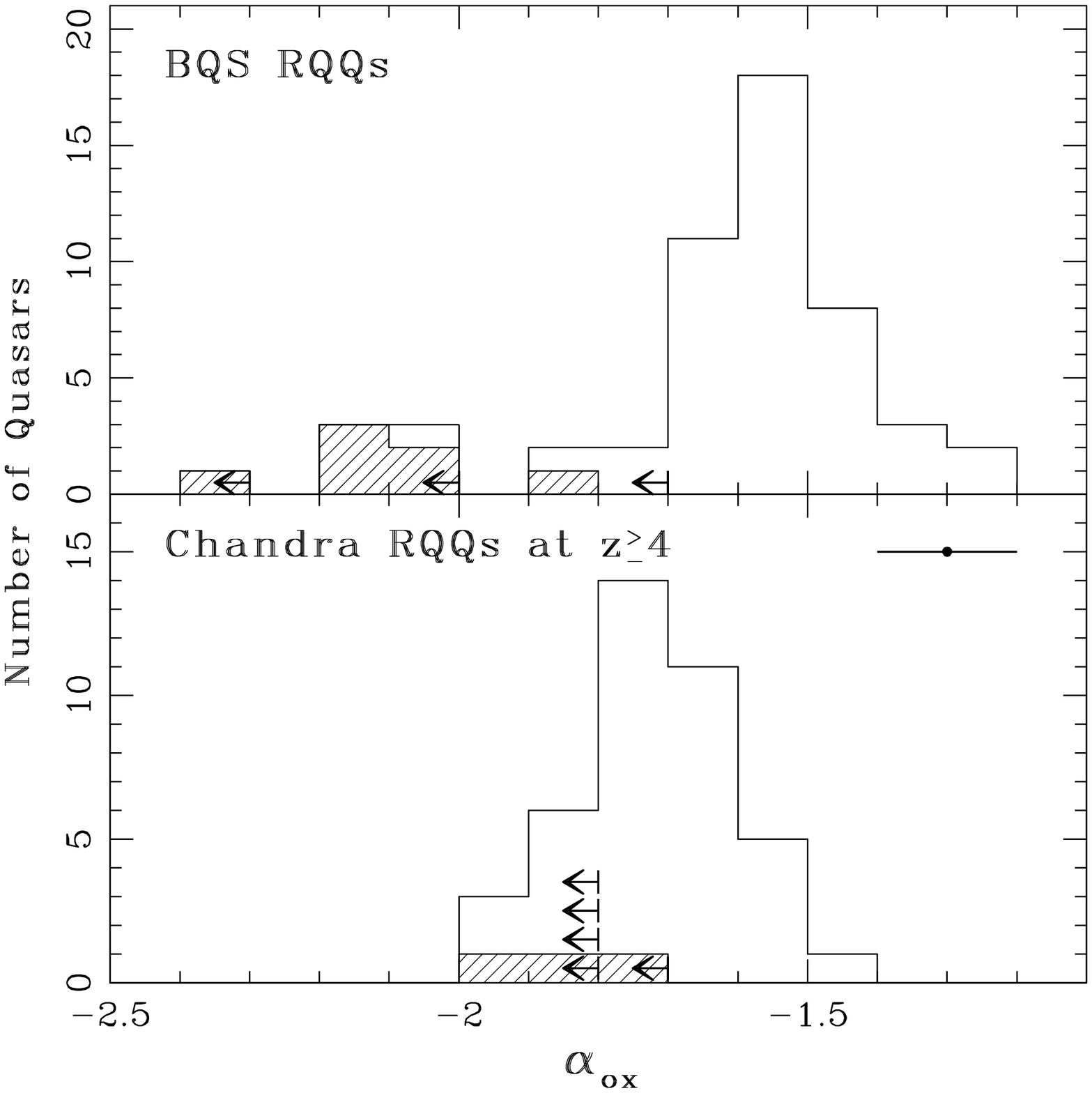}}
\figcaption{\footnotesize 
\aox\ distributions for the BQS RQQs observed by \rosat\ (top) and $z\ge4$ RQQs observed by \chandra\ (bottom). 
The shaded areas indicate the BALQSOs or mini-BALQSOs; 
the areas with arrows indicate 95\% confidence upper limits. 
The average uncertainty in the \aox\ measurements from the \chandra\ observations is shown 
in the bottom panel as the error bar near the upper right. 
\label{fig6}}
\centerline{}
\centerline{}

The principal properties (redshifts, rest-frame 2500~\AA\ and \hbox{2--10~keV} luminosities, and \aox\ values) 
of representative samples of RQQs at low and high redshifts are reported in Table~4. 
%
%

We find no significant correlation of \aox\ with either rest-frame 2500~\AA\ luminosity 
or redshift using only the available $z\ge4$ RQQs. 
However, we can also use the new \xray\ detections and upper limits presented here to extend the 
VBS03 study of \aox\ dependence upon 2500~\AA\ luminosity and redshift.\footnote{See $\S$2.2 of VBS03 
for discussion about the method used to derive 2500~\AA\ fluxes from the SDSS filters.}
As in VBS03, we mostly use the SDSS RQQs taken from the Early Data Release (EDR) quasar catalog 
(Schneider et al. 2002). 
In particular, 
at $z<4$ we use the 128 EDR RQQs of VBS03 (see their $\S$3.1.1) 
covered by pointed \rosat\ PSPC and HRI observations, with an overall \xray\ detection fraction 
of $\approx$~48\%. 
At $z\ge4$ we use all published \chandra\ (V01a; Brandt et al. 2002; Bechtold et al. 2003) 
and \xmm\ (Brandt et al. 2001) observations of 
SDSS RQQs in addition to the ones presented in this paper. 
In total we have 20 SDSS RQQ observations at $z>4$ (including RD~657, observed in one of the fields 
surveyed by the SDSS; see $\S$2.2 and Djorgovski et al. 2003), 16 of which have \xray\ detections. 
To improve the statistics at high luminosities and redshifts without introducing any 
evident bias (see $\S$3.2 of VBS03), 
we also include 15 PSS RQQs and seven BRI RQQs at $z\ge4$ observed by \chandra\ (V01a; 
Bechtold et al. 2003; V03; Priddey et al., in preparation); 
only one of these is not detected in the \xray\ band (V01a). 
%
\scriptsize
\begin{center}
{\sc TABLE 4 \\ Principal Properties of Three Optically Selected Samples \\ of RQQs at Low and High Redshifts}
\vskip 0.3cm
\begin{tabular}{lccc}
\hline
\hline
  & \multicolumn{3}{c}{Sample Name} \\
\cline{2-4}
  & BQS & SDSS & PSS$+$BRI \\ 
$N$ & 46 & 16 & 19 \\ 
$N_{\rm detections}$ & 45 & 13 & 18 \\
$N_{\rm upper\ limits}$ & 1 & 3 & 1 \\ \\
Range of $z$ & 0.061--0.472 & 4.62--6.28 & 4.00--4.73 \\
$\langle\ z\ \rangle$$^{\ \rm a}$ & 0.17$\pm{0.01}$ & 5.11$\pm{0.11}$ & 4.28$\pm{0.04}$ \\
Median $z$ & 0.16 & 4.96 & 4.23 \\ \\
Range of $\log\ l_{\rm 2500~\AA}$ & 29.14--30.70 & 31.22--31.75 & 31.54--32.59 \\
$\langle\ \log\ l_{\rm 2500~\AA}\ \rangle$$^{\ \rm a}$ & 29.66$\pm{0.05}$ & 31.53$\pm{0.04}$ & 31.96$\pm{0.06}$ \\
Median $\log\ l_{\rm 2500~\AA}$ & 29.60 & 31.55 & 31.92 \\ \\
Range of $\log\ L_{\rm 2-10~keV}$ & 43.36--44.79 & 44.53--45.89 & 44.75--45.82 \\ 
$\langle\ \log\ L_{\rm 2-10~keV}\ \rangle$$^{\ \rm a}$ & 44.07$\pm{0.10}$ & 44.99$\pm{0.09}$ & 45.34$\pm{0.07}$ \\ 
Median $\log\ L_{\rm 2-10~keV}$ & 44.13 & 45.03 & 45.41 \\ \\
Range of \aox & $-1.23$ to $-2.06$$^{\ \rm b}$ & $-1.43$ to $-1.82$ & $-1.56$ to $-1.94$ \\
$\langle\ \alpha_{\rm ox}\ \rangle$$^{\ \rm a}$ & $-1.56\pm{0.02}$ & $-1.69\pm{0.03}$ & $-1.73\pm{0.02}$ \\
Median \aox & $-1.55$ & $-1.71$ & $-1.73$ \\ \\
References & (1,2,3,4) & (4,5,6,7,8) & (5,7,9) \\
\hline
\end{tabular}
\vskip 2pt
\parbox{8.5cm}
{\small\baselineskip 9pt
\footnotesize
\indent
{\sc Note. ---} 
The luminosities are in the quasars' rest frames and have been obtained (or converted) 
adopting $H_{0}$=70 km s$^{-1}$ Mpc$^{-1}$ in a $\Lambda$-cosmology 
with $\Omega_{\rm M}$=0.3 and $\Omega_{\Lambda}$=0.7. 
$l_{\rm 2500~\AA}$ is in units of \lumh\ and $L_{\rm 2-10~keV}$ is in units of erg~s$^{-1}$. 
Corrections for the \chandra\ ACIS quantum efficiency decay at low energies have been applied in the computation 
of the \xray\ luminosities and \aox\ values (see $\S$3 for details). \\
BALQSOs and mini-BALQSOs have been excluded from all of the samples used here. 
For the SDSS and PSS$+$BRI samples at $z\ge4$, the values presented here have been obtained 
using \chandra\ and \xmm\ results. 
For the BQS sample, the rest-frame \hbox{2--10~keV} luminosities have been obtained from a representative 
subsample of 21 BQS RQQs observed by \asca\ (George et al. 2000) and \sax\ (Mineo et al. 2000) 
in the observed \hbox{$\approx$~0.5--10~keV} and \hbox{$\approx$~0.1--10~keV} bands, respectively. 
This subsample covers approximately the same range in redshift as the original 
sample of 46 BQS RQQs with $M_{\rm B}<-23$ used in $\S$3.1.1 of Brandt et al. (2001). \\
$^{\rm a}$ The errors on average values represent the 
standard deviation of the mean. 
$^{\rm b}$ \aox=$-2.06$ represents the slope of PG~2214$+$139. 
See Appendix~A of Laor \& Brandt (2002) for a discussion about possible absorption in this quasar. \\
{\sc References. ---} 
(1) Brandt et al. 2000; (2) George et al. 2000; (3) Mineo et al. 2000; 
(4) Brandt et al. 2001; (5) V01a; (6) Brandt et al. 2002; 
(7) Bechtold et al. 2003; (8) this paper; (9) V03.
}
\end{center}
\setcounter{table}{4}
\normalsize
%
Therefore, the final sample used in the present analysis is comprised of 
170 RQQs ($\approx$~58\% with \xray\ detections) over the redshift range \hbox{$z\approx$~0.2--6.3} 
and the rest-frame 2500~\AA\ luminosity range 
\hbox{$\log~(l_{2500~\mbox{\rm \scriptsize\AA}}/{\rm erg~s^{-1}~Hz^{-1}})\approx$~29.0--32.6}. 
Eleven of these quasars are known BALQSOs; 
results obtained both including and excluding BALQSOs are presented below. 
As luminosity is correlated with redshift in flux-limited samples, we use 
a partial correlation analysis technique developed for use with censored data 
(Akritas \& Siebert 1996). 
Partial correlation analyses confirm previous findings (e.g., Avni et al. 1995; VBS03): \aox\ depends 
primarily upon rest-frame 2500~\AA\ luminosity. 
The significance of this anti-correlation is \hbox{3.5--4.0$\sigma$} 
(slightly higher than in VBS03), depending upon whether 
PSS/BRI quasars and BALQSOs are included in the analyses (see Table~5). 

No significant correlation of \aox\ with redshift is found. 
These results suggest that the mechanism driving quasar broad-band emission 
is similar in the local and early Universe, with no evidence 
for unusual phenomena in high-redshift quasar 
nuclei such as ``trapping radius'' effects (e.g., Begelman 1979) 
or accretion-disk instabilities (e.g., Lightman \& Eardley 1974). 
%
\scriptsize
\begin{center}
{\sc TABLE 5 \\ Dependencies of \aox\ upon redshift and 2500~\AA\ luminosity density \\ from partial correlation analysis}
\vskip 0.3cm
\begin{tabular}{lcccc}
\hline
\hline
Sample Name & Number & \aox--$z$ &   & \aox--$\log l_{2500~\mbox{\rm \scriptsize\AA}}$ \\
            & of RQQs &          &   &  \\
SDSS only, with BALQSOs               & 148 & 1.2$\sigma$ & & 3.7$\sigma$ \\
SDSS only, without BALQSOs            & 138 & 0.9$\sigma$ & & 3.5$\sigma$ \\
SDSS $+$ PSS, with BALQSOs            & 163 & 1.3$\sigma$ & & 3.9$\sigma$ \\
SDSS $+$ PSS, without BALQSOs         & 152 & 1.2$\sigma$ & & 3.9$\sigma$ \\
SDSS $+$ PSS $+$ BRI, with BALQSOs    & 170 & 1.4$\sigma$ & & 4.0$\sigma$ \\
SDSS $+$ PSS $+$ BRI, without BALQSOs & 159 & 1.1$\sigma$ & & 4.0$\sigma$ \\
\hline
\end{tabular}
\end{center}
\setcounter{table}{5}
\centerline{}
\centerline{}
\normalsize
%

Although we did not find a significant correlation of \aox\ with 
either rest-frame 2500~\AA\ luminosity or redshift using only the 
$z\ge4$ RQQs, the spread of \aox\ for $z\ge4$ RQQs is unlikely 
to be explained solely by measurement errors. 
To demonstrate this, we have applied the likelihood method described 
in Section~IV of Maccacaro et al. (1988) to calculate the intrinsic 
dispersion of \aox\ (accounting for measurement errors). 
Ignoring upper limits, we find an intrinsic dispersion of 
$\sigma_{\rm i}=0.062^{+0.030}_{-0.033}$ (68\% confidence errors for 
two interesting parameters; see Maccacaro et al. 1988). 
Accounting for upper limits would only increase the intrinsic 
dispersion (see the bottom panel of Fig.~6). For comparison, the 
``straight'' dispersion (not accounting for measurement errors) is 
$\sigma_{\rm s}=0.110$. 

SDSS~0231$-$0728 and RD~657 have the flattest \aox\ values in our sample. 
It is possible (see $\S$4) that they experienced variability over the period between the optical and 
\xray\ observations. An alternative explanation could be the presence of some dust obscuring the optical/ultraviolet 
radiation but not the \xray\ emission. 
The optical spectra of these quasars, however, do not show any clear evidence supporting this scenario. 
Radio emission can affect the quasar broad-band spectral energy distributions 
producing flatter \aox\ values. 
Although significant radio emission can be ruled out for SDSS~0231$-$0728 (see Table~3), the radio-loudness 
upper limit for RD~657 is not sufficiently tight to exclude some radio emission from this quasar. 

SDSS~0913$+$5919, the only radio-loud object in our sample and the highest redshift RLQ with an \xray\ detection, 
has the highest $R$ parameter ($R\approx$~411) and the steepest \aox\ ($-1.67\pm{0.11}$) 
among the RLQs observed thus far by \chandra\ at $z>4$ (V01a; V03). 
\xray\ absorption might be an explanation for the steep \aox\ value, 
although the detection of five of the six source counts in the 
soft band (see Table~2) suggests a different interpretation. 
The anti-correlation between \aox\ and rest-frame 2500~\AA\ luminosity (e.g., Avni et al. 1995; VBS03) 
cannot explain its steep (for a RLQ) \aox\ value, since 
SDSS~0913$+$5919 is also the optically faintest $z>4$ RLQ observed by \chandra\ to date. 

\subsection{Joint X-ray Spectral Fitting of SDSS Quasars at $z>4.8$}

To define the average \xray\ spectral properties of high-redshift SDSS quasars and to make a 
comparison with those obtained for a sample of nine optically luminous PSS quasars 
at \hbox{$z\approx$~4.1--4.5} (V03), 
we have selected a sample of 13 SDSS quasars at $z>4.8$ detected by \chandra\ 
\hbox{($\langle\ z \ \rangle=5.23$)}. 
Their total exposure time is 81.6~ks.  
The quasars in this sample are selected to have more than two counts in the full band; 
this choice excludes SDSS~0756$+$4502 
(presented in this paper) and SDSS~1208$+$0010 (V01a). 
Five of the quasars used in this sample 
have been previously published: the three highest redshift SDSS quasars observed 
thus far (SDSS~1030$+$0524, SDSS~1306$+$0356, and SDSS~0836$+$0054; Brandt et al. 2002); 
SDSS~1204$-$0021 (Bechtold et al. 2003); and SDSS~0211$-$0009 (V01a). 
Eleven of the 13 quasars are radio quiet; the exceptions are SDSS~0836$+$0054 (radio moderate 
with $R\approx$~10) and SDSS~0913$+$5919 (radio loud with $R\approx$~411; see Table~3). 
In the following we present the joint \xray\ spectral analysis both with and without RLQs, 
since RLQs often appear to be characterized by flatter \xray\ photon indices (e.g., Cappi et al. 1997). 
The sample does not appear to be biased by the presence 
of a few high signal-to-noise ratio objects. 
For example, the removal of the radio-moderate quasar 
SDSS~0836$+$0054, the second brightest source in the present sample, does not 
provide significantly different results (see below). 

Source counts were extracted from 2\arcsec\ radius circular apertures 
centered on the \xray\ position of each quasar. The background was taken 
from annuli centered on the quasars, avoiding the presence of nearby \xray\ sources. 
In total we have 120 counts (in the rest-frame \hbox{$\approx$~2.5--36~keV} band) 
from the source extraction regions and five counts from the background 
regions normalized to the size of those used for the sources; half of the background counts are from 
the SDSS~0941$+$5947 field (see $\S$2.1). 
All of the ARFs used in the spectral analysis have been corrected to 
account for the low-energy quantum efficiency decay of ACIS (see $\S$2.1). 
As in $\S$2.1, spectral analysis was carried out with {\sc xspec} using the background-subtracted 
data and {\it C}-statistic (Cash 1979). 
Since {\it C}-statistic fitting is performed using the unbinned data, it retains 
all the spectral information and allows one to associate with each quasar its own Galactic absorption column density 
and redshift (for the fitting with intrinsic absorption). 

Joint spectral fitting with a power-law model (leaving the normalizations free to vary) 
and Galactic absorption provides a good fit to the data as judged by the small data-to-model residuals 
and checked with 10,000 Monte-Carlo simulations. The resulting photon index is 
$\Gamma=1.84^{+0.31}_{-0.30}$. 
Excluding the two non-RQQs 
from the joint spectral fitting gives a similar result ($\Gamma=1.86^{+0.41}_{-0.37}$). 
These photon indices are consistent with those obtained for \hbox{$z\approx$~0--3} samples of RQQs 
in the rest-frame \hbox{2--10~keV} band (e.g., George et al. 2000; Mineo et al. 2000; Reeves \& Turner 2000; 
Page et al. 2003) and the \hbox{$z\approx$~4.1--4.5} 
optically luminous PSS quasars presented in V03 ($\Gamma=1.98\pm{0.16}$). 
As a consistency check, we compared the photon index obtained from joint spectral fitting with that derived from  
band-ratio analysis (i.e., the ratio of the \hbox{2--8~keV} to \hbox{0.5--2~keV} counts; 
we have assumed for the Galactic column density the average value for the 13 quasars weighted by their 
number of counts). In this case 
we obtain $\Gamma=1.85^{+0.22}_{-0.20}$.\footnote{If we increase 
the number of soft-band counts by 10\% to take into account crudely the quantum efficiency decay 
of ACIS at soft energies, we obtain $\Gamma=1.93^{+0.21}_{-0.20}$. 
The small differences in the photon indices obtained via joint spectral fitting and band-ratio analysis 
(with a 10\% increment in the number of soft-band counts) are probably due 
to the fact that only the former retains all the spectral information.} 

We also constrained the presence of neutral intrinsic absorption; 
the joint-fitting technique described above provides a counts-weighted average column density. 
Solar abundances have been assumed, although 
previous optical studies indicate that high-redshift quasar nuclei are often characterized by 
supersolar abundances of heavy elements (e.g., Hamann \& Ferland 1999; Constantin et al. 2002; Dietrich et al. 2003). 
Doubling the abundances in the fit gives a reduction in the column density of a factor of 
$\approx$~2. 
No evidence for absorption in excess of the Galactic column density has been found, 
as shown by the 68, 90, and 99\% confidence regions plotted in Figure~7. 
%
\figurenum{7}
\centerline{\includegraphics[angle=-90,width=8.5cm]{vignali.fig7.ps}}
\vskip 0.7cm
\figcaption{\footnotesize
Confidence regions for the photon index and intrinsic column density 
derived from joint spectral fitting of the 13 SDSS quasars (11 RQQs, two non-RQQs) 
at $z>4.8$ thus far detected by \chandra\ with more than two full-band counts 
(solid contours; see $\S$4.3 for details). 
Dashed contours indicate the results obtained when only the 11 SDSS RQQs 
at $z>4.8$ are considered in the joint spectral fitting. 
\label{fig7}}
\centerline{}
\centerline{}
%
On average, it appears that any neutral \xray\ absorption in $z>4.8$ \xray\ detected SDSS quasars 
has a column density of $N_{\rm H}<3.95\times10^{22}$~cm$^{-2}$ (at the 90\% confidence level; 
$N_{\rm H}<4.37\times10^{22}$~cm$^{-2}$ when only the RQQs are used in the spectral analysis). 
For comparison, \xray\ absorption of \hbox{$\approx$~(2--5)$\times10^{22}$~cm$^{-2}$} 
has been detected in a few RQQs and some RLQs at \hbox{$z\approx$~1--3} (e.g., Cappi et al. 1997; Reeves \& Turner 2000). 
Given the limited counting statistics available for the present sample, it is not possible to obtain useful spectral 
constraints for more complex absorption models (e.g., ionized absorption). 
The typical upper limits on the equivalent widths of narrow iron K$\alpha$ lines (either neutral or ionized) are 
in the range 420--950~eV. 

The \xray\ spectral results at $z>4$ presented here and in V03 allow us to place constraints on 
any spectral changes in the \xray\ continua of RQQs as a function of redshift. 
In Figure~8 we plot the photon index as a function of redshift for some optically selected samples of RQQs. 
The photon indices plotted in the figure are for the nuclear, high-energy ($\simgt2$~keV) component, in order 
to avoid spectral complexities such as soft excesses. 
The \asca\ RQQs have been taken from Vignali et al. (1999; V99 in the key), 
George et al. (2000; G00), Reeves \& Turner (2000; RT00), and Vignali et al. (2001b; V01b). 
The \sax\ RQQs are from Mineo et al. (2000; M00). 
At the highest redshifts we plot the $z=3.91$ gravitationally lensed BALQSO APM~08279$+$5255 (Chartas et al. 2002; C02), 
the results obtained from joint \xray\ spectral fitting of nine PSS quasars in the redshift range 
\hbox{$z\approx$~4.1--4.5} (V03), 
and the 11 SDSS RQQs at $z>4.8$ presented in this paper. 
The distribution of \xray\ photon indices is characterized by significant scatter, 
so precise parametric modeling of any $\Gamma$ versus redshift relation is difficult. 
However, according to a Spearman's $\rho$ correlation test, a correlation 
between $\Gamma$ and redshift is not significant (2.3$\sigma$). 
Applying a linear regression algorithm that accounts for intrinsic scatter and measurement errors 
(Akritas \& Bershady 1996), we obtain $\mid \frac{\partial \ \Gamma}{\partial \ z} \mid$ $<0.05\pm{0.03}$. 
%
\figurenum{8}
\centerline{\includegraphics[angle=0,width=8.5cm]{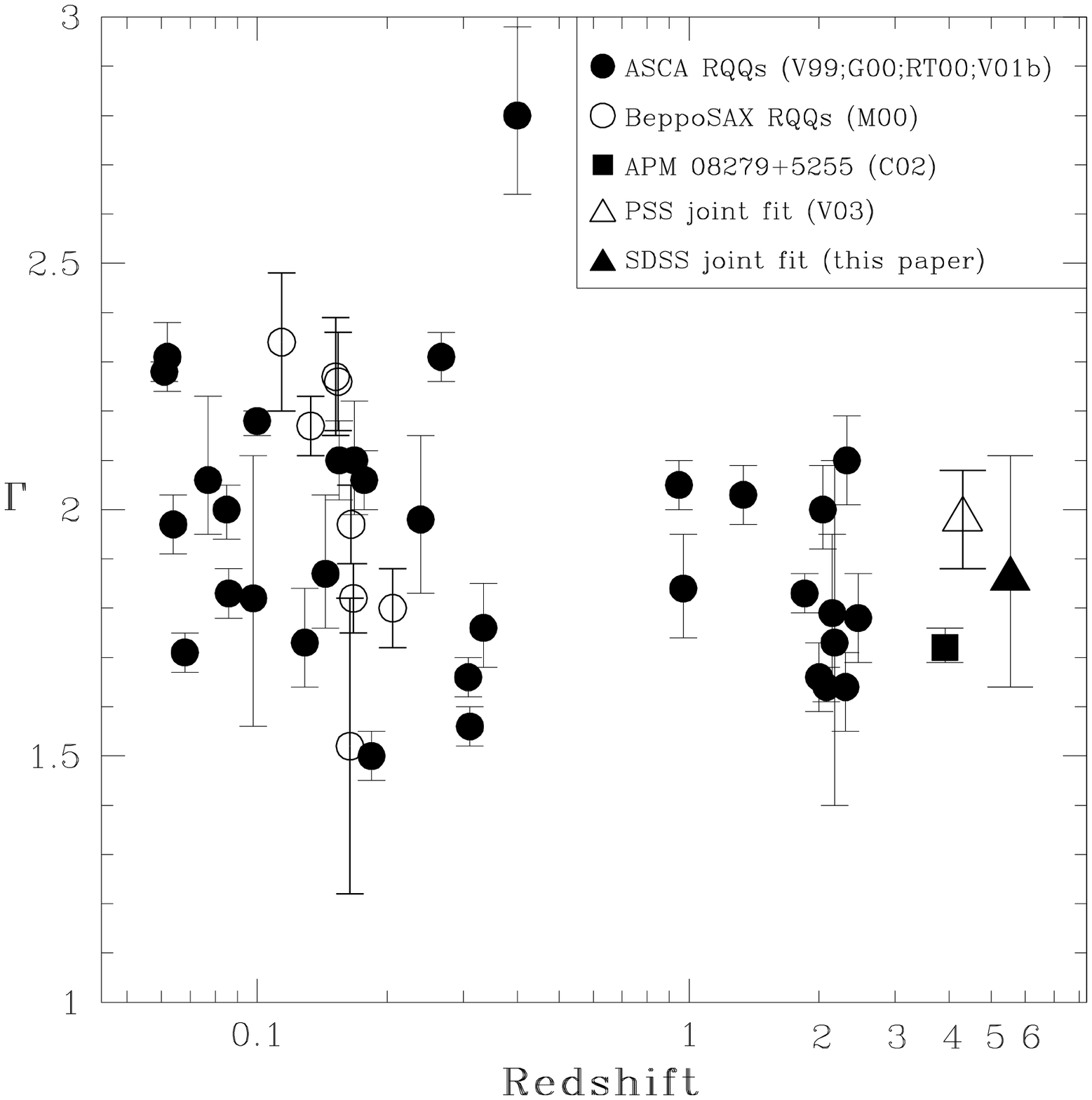}}
\figcaption{\footnotesize 
Plot of photon index versus redshift for optically 
selected RQQs. 
The \asca\ RQQs have been taken from Vignali et al. (1999; V99 in the key), 
George et al. (2000; G00), Reeves \& Turner (2000; RT00), and Vignali et al. (2001b; V01b). 
The \sax\ RQQs are from Mineo et al. (2000; M00). 
At the highest redshifts we plot the $z=3.91$ gravitationally lensed 
BALQSO APM~08279$+$5255 (Chartas et al. 2002; C02), 
the results obtained from joint \xray\ spectral fitting of nine PSS quasars 
in the redshift range $z\approx$~4.1--4.5 (V03), 
and the 11 SDSS RQQs at $z>4.8$ presented in this paper (the latter two data points are plotted at the 
average redshifts of the relevant papers). 
Error bars are at the 68\% confidence level for all data points.  
The object with the steepest \xray\ photon index in this plot, PG~1543$+$489 (George et al. 2000), 
is known to be a narrow-line type~1 quasar (Boroson \& Green 1992). 
\label{fig8}}
\centerline{}
\centerline{}
%

\section{Summary}

We have reported the \chandra\ and \xmm\ observations of a sample of 13 quasars 
at \hbox{$z\approx$~4.7--5.4}, the majority of them selected by the SDSS, and compared their properties 
with those of all the $z\ge4$ quasars previously observed in the \xray\ band. 
The principal results are the following: \\
\begin{itemize}
\item 
Eleven quasars have been detected in the \xray\ band, 
doubling the number of $z\ge4.8$ \xray\ detected quasars. 
The \xray\ detections include SDSS~0913$+$5919, 
the highest redshift ($z=5.11$) RLQ thus far detected in the \xray\ band. 
Two quasars, SDSS~0338$+$0021 and SDSS~1737$+$5828, were not detected by the present observations. 
The former non-detection is probably due to the short exposure time of the \xmm\ observation after removal 
of flaring-background intervals. 
For the latter quasar, strong \xray\ absorption 
(\hbox{$N_{\rm H}\simgt1.2\times10^{23}$~cm$^{-2}$} at the source redshift) is a likely explanation. 
An HET spectrum of this source, revealing absorption features, seems to support this hypothesis. 
For our sample, no evidence for source extension or an overdensity of companion objects has been found. 
\item
The presence of a significant correlation between the soft \xray\ flux and \ab1450\ magnitude has been 
extended to the highest redshifts. This indicates that the same engine 
(namely accretion onto a supermassive black hole) is powering the ultraviolet and \xray\ emission. 
\item 
The anti-correlation between \aox\ and rest-frame 2500~\AA\ luminosity density found for SDSS RQQs 
over the redshift range \hbox{$z\approx$~0.2--6.3} (VBS03) 
has been confirmed (\hbox{3.5--4.0$\sigma$} significance level) 
after the inclusion of the RQQs presented in this paper. 
Similarly to VBS03, no significant correlation between \aox\ and redshift has been found. 
These results suggest that the mechanism driving quasar broad-band emission 
is similar in the local and early Universe, with no evidence for unusual phenomena in high-redshift quasar 
nuclei such as ``trapping radius'' effects or accretion-disk instabilities. 
\item 
The joint \hbox{$\approx$~2.5--36~keV} rest-frame \xray\ spectrum of the 13 
SDSS quasars at $z>4.8$ observed thus far by \chandra\ 
is well parameterized by a simple power-law model 
with $\Gamma=1.84^{+0.31}_{-0.30}$ and no evidence for intrinsic absorption 
($N_{\rm H}\simlt4.0\times10^{22}$~cm$^{-2}$ at 90\% confidence). 
\item 
There is no evidence for significant ($>30\%$) optical variability over the 
time interval of a few years (in the observed frame) between the SDSS and HET observations. 
\end{itemize}

Among the quasars presented in this paper, the only object suitable for moderate-quality ($\approx$~1000 photons) 
\xray\ spectroscopy in a 50~ks \xmm\ observation is SDSS~0231$-$0728. 
\xray\ spectroscopy of the other quasars will be possible only with future missions such as 
\conx, \xeus, and \genx\ (see V03 for some relevant simulations).

\acknowledgments

We gratefully acknowledge the financial support of NASA LTSA grant NAG5-8107 (CV, WNB), 
\chandra\ \xray\ Center grant GO2-3134X (CV, WNB, DPS), NASA grant NAG5-9918 (CV, WNB), 
and NSF grant AST99-00703 (DPS). 
CV also acknowledges partial support from the Italian Space Agency under contract 
ASI I/R/113/01 and I/R/073/01. 
The authors would like to thank the referee P. Hewett for his comments, 
T.~Abel for useful suggestions, 
D.~Alexander for thoughtful comments and help with Monte-Carlo simulations, 
A.~Baldi for help with \xmm\ data reduction, 
F.~Bauer for help with IDL codes, 
G.~Brunetti and G. Chartas for useful discussions,
S.~Djorgovski for providing us with the results on RD~657 before their publication, 
G.~Garmire for discussions about ACIS quantum efficiency decay, and 
L.~Ramsey for kindly allocating the time for the HET spectroscopic observation presented here. 

The HET is a joint project of the University of Texas 
at Austin, the Pennsylvania State University, Stanford University, 
Ludwig-Maximillians-Universit\"at M\"unchen, and Georg-August-Universit\"at G\"ottingen. 
The HET is named in honor of its principal benefactors, 
William P. Hobby and Robert E. Eberly. The Marcario Low-Resolution 
Spectrograph is named for Mike Marcario of High Lonesome Optics, who 
fabricated several optics for the instrument but died before its completion; 
it is a joint project of the HET partnership and the 
Instituto de Astronom\'{\i}a de la Universidad Nacional Aut\'onoma de M\'exico.

Funding for the creation and distribution of the SDSS Archive has been provided by 
the Alfred P. Sloan Foundation, the Participating Institutions, the National Aeronautics 
and Space Administration, the National Science Foundation, the U.S. Department of Energy, 
the Japanese Monbukagakusho, and the Max Planck Society. 
The SDSS Web site is http://www.sdss.org/. 
The SDSS is managed by the Astrophysical Research Consortium (ARC) for the Participating Institutions. 
The Participating Institutions are the University of Chicago, Fermilab, the Institute for Advanced Study, 
the Japan Participation Group, the Johns Hopkins University, Los Alamos National Laboratory, 
the Max-Planck-Institute for Astronomy (MPIA), the Max-Planck-Institute for Astrophysics (MPA), 
New Mexico State University, the University of Pittsburgh, Princeton University, the United States Naval Observatory, 
and the University of Washington.

\clearpage

\appendix

Here we report the updated \xray\ flux values (and derived quantities) 
for $z\ge4$ quasars observed by \chandra\ in the past. 
The \xray\ values have been corrected for the ACIS quantum energy decay at low energies. 
%
\vglue -3cm
\begin{deluxetable}{lccccccccccc}
\tablenum{A1}
\tablecolumns{12}
\tabletypesize{\footnotesize}
\tablewidth{0pt}
\tablecaption{Optical, X-ray, and Radio Properties of $z\ge4$ Quasars Previously Observed by \chandra}
\tablehead{ 
\colhead{Object} & \colhead{$z$} & \colhead{\ab1450} & \colhead{$f_{\rm 2500~\AA}$} & 
\colhead{$\log (L_{\rm 2500~\AA})$\tablenotemark{a}} & \colhead{$M_B$} & 
\colhead{$f_{\rm 0.5-2~keV}$} & \colhead{$f_{\rm 2~keV}$} & 
\colhead{$\log (L_{\rm 2~keV})$\tablenotemark{b}} & \colhead{$\log (L_{\rm 2-10~keV})$} & 
\colhead{\aox} & \colhead{$R$}
}
\startdata
PSS~0059$+$0003  & 4.16 & 19.3 &  {\phn}1.06 & 31.54 & $-$27.62 &      10.2  & {\phn}7.88  &  27.41   &  45.30   &  $-$1.59 & $<3.3$ \\
BR~0103$+$0032   & 4.43 & 18.7 &  {\phn}1.85 & 31.82 & $-$28.33 &      20.1  &      16.28  &  27.77   &  45.66   &  $-$1.56 & {\phn}{\phn}2.9   \\
PSS~0121$+$0347  & 4.13 & 18.5 &  {\phn}2.22 & 31.86 & $-$28.41 &      41.3  &      31.65  &  28.01   &  45.90   &  $-$1.48 & 300.3   \\
PSS~0133$+$0400  & 4.15 & 18.2 &  {\phn}2.93 & 31.98 & $-$28.72 &      20.8  &      16.01  &  27.72   &  45.61   &  $-$1.64 & $<4.4$ \\
PSS~0134$+$3307  & 4.53 & 18.5 &  {\phn}2.22 & 31.92 & $-$28.57 &      11.5  & {\phn}9.45  &  27.55   &  45.44   &  $-$1.68 & $<6.1$ \\
PSS~0209$+$0517  & 4.14 & 17.8 &  {\phn}4.24 & 32.14 & $-$29.12 &      15.2  &      11.66  &  27.58   &  45.47   &  $-$1.75 & $<3.0$ \\
SDSS~0210$-$0018 & 4.77 & 19.3 &  {\phn}1.06 & 31.63 & $-$27.85 &      37.3  &      32.12  &  28.11   &  46.00   &  $-$1.35 & {\phn}86.1 \\
SDSS~0211$-$0009 & 4.90 & 20.0 &  {\phn}0.56 & 31.37 & $-$27.20 & {\phn}3.8  & {\phn}3.36  &  27.15   &  45.04   &  $-$1.62 & $<7.5$ \\
BR~0241$-$0146   & 4.06 & 18.4 &  {\phn}2.44 & 31.88 & $-$28.48 & {\phn}4.3  & {\phn}3.26  &  27.01   &  44.84   &  $-$1.87 & $<1.5$ \\
PSS~0248$+$1802  & 4.43 & 18.1 &  {\phn}3.21 & 32.06 & $-$28.93 &      21.2  &      17.16  &  27.79   &  45.68   &  $-$1.64 & $<4.2$ \\
BR~0308$-$1734   & 4.00 & 18.1 &  {\phn}3.21 & 31.99 & $-$28.76 & {\phn}4.1  & {\phn}3.06  &  26.97   &  44.86   &  $-$1.93 & $<3.9$ \\
BR~0401$-$1711   & 4.23 & 18.9 &  {\phn}1.54 & 31.71 & $-$28.05 &      12.9  &      10.07  &  27.53   &  45.42   &  $-$1.61 & $<8.5$ \\
SDSS~0836$+$0054 & 5.82 & 18.8 &  {\phn}1.67 & 31.96 & $-$28.67 &      12.3  &      12.50  &  27.83   &  45.73   &  $-$1.58 & {\phn}{\phn}9.8   \\
PSS~0926$+$3055  & 4.19 & 16.7 &       11.67 & 32.59 & $-$30.24 &      33.5  &      25.94  &  27.93   &  45.82   &  $-$1.79 & $<0.3$ \\
PSS~0955$+$5940  & 4.34 & 18.4 &  {\phn}2.44 & 31.93 & $-$28.60 & {\phn}6.0  & {\phn}4.79  &  27.22   &  45.12   &  $-$1.81 & $<5.4$ \\
PSS~0957$+$3308  & 4.20 & 18.2 &  {\phn}2.93 & 31.99 & $-$28.74 &      11.3  & {\phn}8.78  &  27.46   &  45.36   &  $-$1.74 & $<1.1$ \\
SDSS~1030$+$0524 & 6.28 & 19.7 &  {\phn}0.76 & 31.67 & $-$27.94 & {\phn}2.5  & {\phn}2.73  &  27.22   &  45.12   &  $-$1.71 & $<6.8$ \\
BR~1033$-$0327   & 4.51 & 18.8 &  {\phn}1.69 & 31.80 & $-$28.26 & {\phn}8.4  & {\phn}6.87  &  27.41   &  45.30   &  $-$1.69 & $<2.5$ \\
PSS~1057$+$4555  & 4.12 & 17.6 &  {\phn}5.09 & 32.22 & $-$29.31 &      25.3  &      19.34  &  27.80   &  45.68   &  $-$1.70 & {\phn}{\phn}2.3   \\
SDSS~1129$-$0142 & 4.85 & 19.2 &  {\phn}1.17 & 31.68 & $-$27.98 &     $<2.4$ &    $<1.75$  & $<26.86$ & $<44.84$ & $<-1.85$ & $<3.4$ \\
SDSS~1204$-$0021 & 5.03 & 19.1 &  {\phn}1.28 & 31.75 & $-$28.14 & {\phn}4.6  & {\phn}4.13  &  27.26   &  45.15   &  $-$1.72 & $<3.4$ \\
SDSS~1208$+$0010 & 5.27 & 20.5 &  {\phn}0.35 & 31.22 & $-$26.82 & {\phn}1.3  & {\phn}1.24  &  26.77   &  44.66   &  $-$1.71 & $<13.1$ \\
PC~1247$+$3406   & 4.90 & 19.2 &  {\phn}1.17 & 31.69 & $-$28.00 &      11.0  & {\phn}9.66  &  27.61   &  45.50   &  $-$1.57 & $<3.3$ \\
SDSS~1306$+$0356 & 5.99 & 19.6 &  {\phn}0.85 & 31.68 & $-$27.97 & {\phn}5.7  & {\phn}5.94  &  27.53   &  45.42   &  $-$1.59 & $<5.3$ \\
PSS~1317$+$3531  & 4.36 & 18.9 &  {\phn}1.54 & 31.73 & $-$28.10 & {\phn}3.6  & {\phn}2.91  &  27.01   &  44.90   &  $-$1.81 & $<2.3$ \\
PSS~1326$+$0743  & 4.09 & 17.2 &  {\phn}7.36 & 32.37 & $-$29.70 &      29.4  &      22.32  &  27.85   &  45.74   &  $-$1.73 & $<0.5$ \\
PSS~1347$+$4956  & 4.51 & 17.4 &  {\phn}6.12 & 32.36 & $-$29.66 &      17.5  &      14.35  &  27.73   &  45.62   &  $-$1.78 & {\phn}{\phn}0.1   \\
PSS~1435$+$3057  & 4.35 & 19.1 &  {\phn}1.28 & 31.65 & $-$27.90 &     $<3.6$ &    $<2.87$  & $<27.00$ & $<44.90$ & $<-1.79$ & $<2.9$ \\
PSS~1443$+$2724  & 4.42 & 19.0 &  {\phn}1.40 & 31.70 & $-$28.03 &      11.3  & {\phn}9.10  &  27.51   &  45.41   &  $-$1.61 & $<2.8$ \\
PSS~1443$+$5856  & 4.26 & 17.7 &  {\phn}4.64 & 32.20 & $-$29.26 & {\phn}4.7  & {\phn}3.66  &  27.09   &  44.99   &  $-$1.96 & $<2.8$ \\
SDSS~1532$-$0039 & 4.62 & 19.4 &  {\phn}0.97 & 31.57 & $-$27.70 &     $<2.5$ &    $<2.10$  & $<26.91$ & $<44.80$ & $<-1.79$ & $<4.3$ \\
SDSS~1605$-$0112 & 4.92 & 19.4 &  {\phn}0.97 & 31.61 & $-$27.80 &     $<3.0$ &    $<2.62$  & $<27.05$ & $<44.94$ & $<-1.75$ & $<4.5$ \\
BR~2212$-$1626   & 4.00 & 18.6 &  {\phn}2.03 & 31.79 & $-$28.25 &      13.4  & {\phn}9.95  &  27.48   &  45.38   &  $-$1.65 & $<6.2$ \\
\tableline
\enddata
\tablecomments{Units are the same as in Table~3.} 
\tablenotetext{a}{Rest-frame 2500~\AA\ luminosity density (\lumh).}
\tablenotetext{b}{Rest-frame 2~keV luminosity density (\lumh).}
\label{app1}
\end{deluxetable}

\end{document}